# Plasmon-driven substitution of 4-mercaptophenylboronic acid to 4-nitrothiophenol monitored by surface-enhanced Raman spectroscopy

Jan Kozisek[a], Jana Hrncirova[a], Miroslav Slouf[b] , and Ivana Sloufova*[a]

[a]*Charles University, Faculty of Science, Department of Physical and Macromolecular Chemistry, Hlavova 2030, 128 40 Prague 2, Czech Republic, E-mail: ivana.sloufova@natur.cuni.cz.*
[b]*Institute of Macromolecular Chemistry, Czech Academy of Sciences, Heyrovskeho nam. 2, 162 06 Prague 6, Czech Republic.*

## Abstract

Plasmon-driven reactions on plasmonic nanoparticles (NPs) occur under significantly different conditions from those of classical organic synthesis and provide a promising pathway for enhancing the efficiency of various chemical processes. However, these reactions can also have undesirable effects, such as 4-mercaptophenylboronic acid (MPBA) deboronation. MPBA chemisorbs well to Ag NPs through its thiol group and can subsequently bind to diols, enabling the detection of various biological structures by surface-enhanced Raman scattering (SERS), but not upon its deboronation. To avoid this reaction, we investigated the experimental conditions of MPBA deboronation on Ag NPs by SERS. Our results showed that the level of deboronation strongly depends on both the morphology of the system and the excitation laser wavelength and power. In addition, we detected not only the expected products, namely thiophenol and biphenyl-4,4-dithiol, but also 4-nitrothiophenol (NTP). The crucial reagent for NTP formation was an oxidation product of hydroxylamine hydrochloride, the reduction agent used in Ag NP synthesis. Ultimately, this reaction was replicated by adding $NaNO_2$ to the system, and its progress was monitored as a function of the laser power, thereby identifying a new reaction of plasmon-driven $-B(OH)_2$ substitution for $-NO_2$.

## Highlights

- Formation of 4-nitrobenzenethiol from 4-mercaptophenyl boronic acid
- Products of 4-mercaptophenyl boronic acid deboronation monitored by SERS
- Experimental conditions affect reactions of 4-mercaptophenyl boronic acid

## 1 Introduction

Plasmonic metal nanoparticles (NPs) or nanostructures, especially Au, Ag and Cu NPs, are efficient light scatters and absorbers [1,2] widely used as substrates in many plasmon-enhanced spectroscopy studies [3–6]. When NPs are irradiated at a suitable excitation wavelength, collective oscillations of free electrons occur at the same frequency as that of the incident excitation radiation. These oscillations are known as localized surface plasmon resonance (LSPR) [1,2].

LSPR excitation markedly enhances the electromagnetic field near plasmonic metal nanostructures. This enhancement strongly depends on the morphology of the nanostructures and is not distributed uniformly. The largest enhancements are confined to regions known as "*hot spots*" [7–9]. One of the most widely used spectroscopic techniques involving LSPR excitation is surface-enhanced Raman scattering (SERS) [10–14]. In SERS, both incident and Raman scattered photons of molecules near plasmonic nanostructures can be enhanced. For this reason, SERS is often used as a highly efficient analytical method for detecting substances at nanomolar concentrations or even at the single-molecule level [8,15].

Beyond plasmon-enhanced spectroscopy, LSPR excitation can also be applied in several plasmon-catalyzed or plasmon-driven reactions [16–21]. In metal NPs, the chemical reaction can be catalyzed by LSPR due to optical near-field enhancement, heat generation, or hot-charge carrier effects [16,18,20,22,23]. Accordingly, a wide range of plasmon-catalyzed (or plasmon-mediated) [17,24] or have already been described, including selective reductions [25–28] and oxidations [29], cross-coupling reactions [30,31], additions and eliminations [32].

Among such reactions involving aromatic thiol compounds, research efforts have focused on plasmon-catalyzed 4-aminobenzenethiol dimerization to 4,4'-dimercaptoazobenzene [33–41] and 4-aminothiophenol oxidation to 4-nitrothiophenol (NTP) and *vice versa*[26,42–46], on 4-mercaptobenzoic acid decarboxylation [47], and on 4-mercaptophenylboronic acid (MPBA) deboronation [48]. MPBA deboronation is an undesirable reaction, though, because MPBA is a suitable linker for molecules with low affinity to plasmonic NP surfaces. Under normal conditions, MPBA enables their detection by SERS [49–52] thanks to its strong affinity for Ag and Au surfaces via its SH group [53]. However, MPBA deboronation prevents MPBA from serving as a linker. In fact, boronic acids such as MPBA have long been widely used in organic chemistry [54] for their ability to participate in coupling reactions, by forming a C-C bond between two aromatic nuclei, and to directly react with diols via their -$B(OH)_2$ group. So as long as MPBA deboronation is avoided, MPBA may be used in analytical or pharmaceutical chemistry to detect a wide range of substances using various techniques, including SERS [50,51,55].

Yet, despite the wide range of potential MPBA applications, laser irradiation results in MPBA deboronation to thiophenol (TP, benzenethiol) [48,56], as in the decarboxylation [57] of 4-mercaptobenzoic acid or in the dehalogenation [58,59] of halogenated thiophenols. The formation of TP on the surface of plasmonic NPs is expressed in SERS spectra as a characteristic doublet [60–65] at ca 998 and 1020 $cm^{-1}$ typical of mono-, but not disubstituted aromatic thiols [66]. Nevertheless, these bands in SERS spectra have often been misassigned [67–72] (i.e., some peaks have been misattributed to MPBA). Alternatively, the detection of a substance by SERS spectroscopy has been based on indirect evidence when authors disregard or overlook deboronation during their data interpretation [73–81]. Thus, describing the conditions under which MPBA deboronation occurs is crucial for all studies in which this molecule is used as a reactive linker to detect other molecules or as a reactant, e.g., in coupling reactions.

Considering the above, the present study initially aimed at describing the conditions of MPBA deboronation in SERS systems in terms of experimental setup, morphology of SERS active systems, excitation wavelength and laser power selections. However, during our experiments, we found that 4-nitrothiophenol (NTP) is also formed in this process in addition to the previously known reactions of MPBA on surfaces – i.e., the aforementioned MPBA deboronation, which leads to TP formation, and the homocoupling reaction upon the formation of biphenyl-4,4′-dithiol (BPDT). Reactions of boronic acids yielding aromatic nitro compounds are known, but under completely different conditions. Typically, NTP has been synthesized from MPBA in non-aqueous solvents and at high temperatures (>70°C) [82–85]. But as found in this study, this reaction can also occur on Ag NPs surfaces in aqueous conditions at room temperature. Therefore, we also aimed at identifying the conditions under which this new plasmon-driven reaction can spontaneously occur in Ag NPs systems. Our results demonstrate that the crucial reagent is an oxidation product of hydroxylamine hydrochloride, a reduction agent commonly used in Ag NPs synthesis, and that this substitution reaction of MPBA to NTP can be replicated by adding $NaNO_2$ to the system while monitoring the progress of this reaction as a function of time and laser power.

# 2 Experimental

## 2.1 Materials

The chemicals used in this study: silver nitrate ($AgNO_3$, crystalline, extra pure), trisodium citrate dihydrate ($HOC(COONa)(CH_2COONa)_2{\cdot}2H_2O$, >99%), sodium borohydride ($NaBH_4$, p.a.), ethanol ($CH_3CH_2OH$, UVASOL), and dichloromethane ($CH_2Cl_2$, >99.8%) were purchased from Merck Life Science; hydroxylamine hydrochloride ($NH_2OH.HCl$, 99.999%), tetrachloroauric(III) acid trihydrate ($HAuCl_4{\cdot}3H_2O$, 99.9%), potassium nitrate ($KNO_3$, >99%), and sodium nitrite ($NaNO_2$, >97%) were purchased from Sigma-Aldrich; thiophenol (**Scheme 1**, TP, benzenethiol, >98%) was purchased from Fluka Chemika; 4-mercaptophenylboronic acid (**Scheme 1**, MPBA, 95%), 4-nitrothiophenol (**Scheme 1**, NTP, >96%) and biphenyl-4,4′-dithiol (**Scheme 1**, BPDT, 95%) was purchased from abcr GmbH; and sodium hydroxide (NaOH, p.a.) was purchased from Lachema. Doubly distilled deionized water was used to prepare all samples.

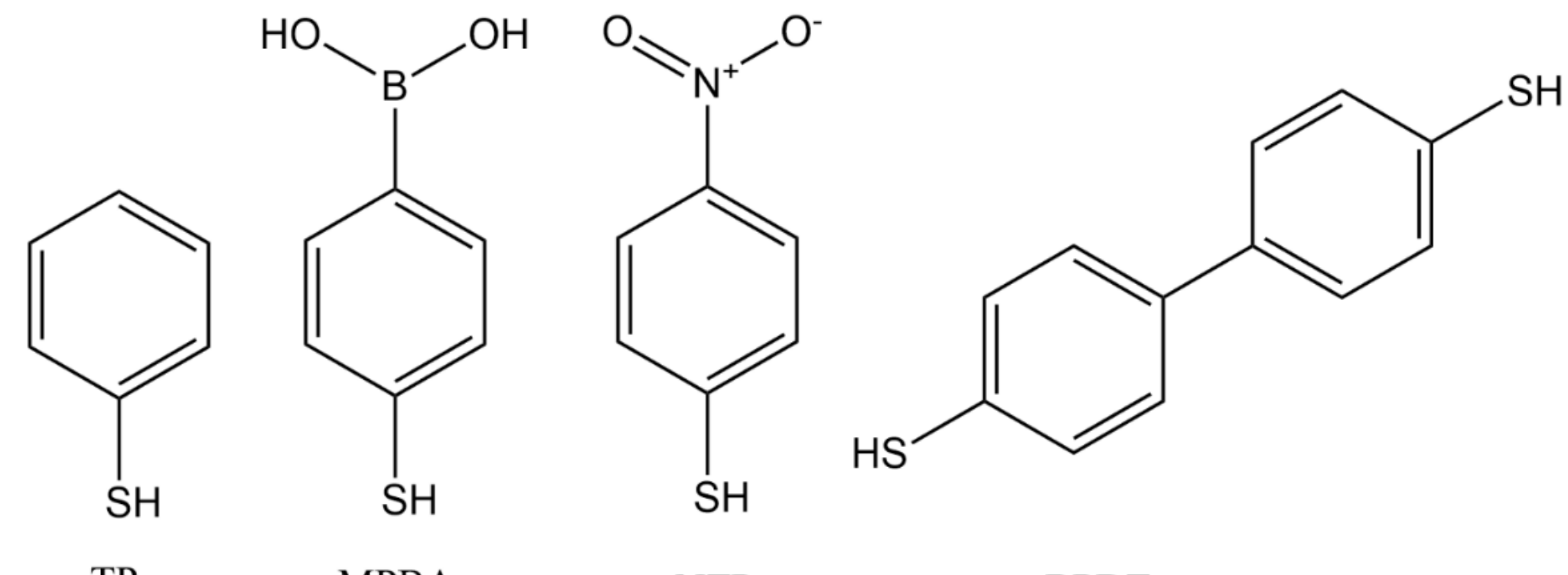

**Scheme 1** – Structures of molecules used in this study: TP – thiophenol, MPBA – 4-mercaptophenylboronic acid, NTP – 4-nitrothiophenol, BPDT - biphenyl-4,4′-dithiol.

## 2.2 Nanoparticle synthesis

Several types of NPs (nanoparticle) hydrosols were synthesized according to the following preparation procedures: **Ag(B) NPs** averaging 10 nm in diameter (**Figure S1**) and with maximum localized surface-plasmon resonances (LSPR) at 392 nm ($\lambda_{max}$= 392 nm; **Figure S2**) were prepared by silver nitrate reduction with sodium borohydride in aqueous environment using the procedure described in ref.[86]; **Ag(N) NPs** averaging 36 nm in diameter (**Figure S1**) and with $\lambda_{max}$= 400 nm (**Figure S2**) were prepared by silver nitrate reduction with hydroxylamine hydrochloride according to the procedure described by Leopold and Lendl [87]. **Ag(C) NPs** with a broad average diameter distribution between 15 and 200 nm (**Figure S1**) and with $\lambda_{max}$= 403 nm (**Figure S2**) were prepared according to a modified procedure by Lee and Meisel [88]: Initially, $AgNO_3$ (18 mg) was dissolved in water (10 mL) and added to 90 mL of boiling water; then, when the solution of $AgNO_3$ started to boil, a solution of 1% sodium citrate (2 mL) was added, and this solution was subsequently heated to boiling for another 90 minutes. **Au(N) NPs** averaging 75 nm in diameter (**Figure S1**) were prepared according to modified procedure for preparation Ag(N) NPs by Leopold and Lendl [87]: Initially, to hydroxylamine hydrochloride (10.4 mg) dissolved in water (90 mL) was added 0.3 mL of 1 M solution of NaOH. Then, under vigorous stirring $HAuCl_4.3H_2O$ ($5\times10^{-3}$ M, 5 mL) was added and the final solution was continuously stirred for 45 min. **Au(C) NPs** averaging 22 nm in diameter (**Figure S1**) and with $\lambda_{max}$= 522 nm (**Figure S2**) were prepared by $HAuCl_4$ reduction with sodium citrate according to the procedure described by Lee and Meisel [88]. Bimetallic **AgAu(N) NPs** averaging 15 nm in diameter (**Figure S1**) and with $\lambda_{max}$= 532 nm (**Figure S2**) were prepared according to the procedure described by Srnova-Sloufova et al. [89]: 12 mL of Ag(B) NPs were mixed with 30 mL of water; then, hydroxylamine hydrochloride ($6.25\times10^{-3}$ M, 25 mL) and $HAuCl_4$ ($4.65\times10^{-4}$ M, 25 mL) solutions were simultaneously added under vigorous stirring to the diluted solution of Ag(B) NPs, continuously stirring the solution for another 45 min.

## 2.3 Preparation of SERS-Active systems

**Aggregated systems in cuvettes:** MPBA, TP and NTP were diluted in ethanol to obtain 16 mM solutions. Unless otherwise directly specified in the text, all SERS active systems were prepared by mixing 2 mL of NPs solution with 20 µL of a 16 mM solution of the appropriate adsorbate. Henceforth, all systems prepared in this study are referred to as, for example, the Ag(N)-MPBA system (i.e. system prepared by mixing 2 mL of Ag(N) NPs with 20 µL of 16 mM ethanolic solution of MPBA). Preparing system with NTP required adding 100 µL of a 1 M NaCl aqueous solution to the hydrosol due to weak aggregation and to the inability to acquire SERS spectra at 785 nm excitation.

**2-dimensional (2D) arrays of NPs.** These 2D arrays were prepared according to a modified procedure reported earlier [90]. Briefly, 2.5 ml of Ag or Au hydrosol and 20 µL of a 16 mM solution of MPBA and 2.5 ml of dichloromethane was vigorously shaken for several minutes. After separation of the aqueous and dichloromethane phases, the nanoparticles self-assembled on the interphase. The layer of NPs was then collected and transferred onto a glass slide where the fraction of NPs spontaneously reassembled into a layer of NPs on a water drop. The excess liquid was slowly drained using a filter paper.

## 2.4 Instrumentation

**SERS spectra** excited at 445 nm and 633 nm (diode lasers) were measured on a DXR Raman spectrometer (Thermo Scientific). A 50x objective was used for solid samples; a macro-adapter was used for liquid samples. A software-controlled spectrometer calibration procedure was used. The laser power at the sample was set to 8-10 mW for liquid samples and 0.1 mW for solid samples. The full-scale grating (200-3200 cm$^{-1}$ spectral range) was used in all measurements. Time series of *SERS* spectra excited at 532 and 785 nm (diode lasers) were collected on a MonoVista CRS+ Raman system (Spectroscopy & Imaging GmbH, Germany) equipped by 150 lines/mm grating (for 532 nm excitation used in second order) and spectrograph with 750 mm focal distance. The laser power at the sample was 0.1-11 mW laser. Spectrometer wavelength and intensity were calibrated using the software-controlled auto alignment procedure with mercury and Ne-Ar lamps.

**Surface-plasmon extinction (SPE) spectra** were measured on a Shimadzu UV-2401 spectrophotometer using quartz cuvettes (with 1.0 and 0.2 cm optical paths).

**Transmission electron microscopy (TEM)** analysis of the NPs was performed under a Tecnai G2 Spirit Twin 12 TEM microscope (FEI; Czech Republic). The samples were observed in bright field imaging mode at 120 kV accelerating voltage.

### 2.5 Processing of spectral sets

**Baseline correction and analysis of time-evolution SERS spectral sets** was made using singular value decomposition and factor analysis (FA) methods as described in Text S1 in supplementary material according to Palacky et al [91].

**Calculation of the SERS excitation profile of TP**. The sample used to determine the SERS excitation profile of TP was prepared by mixing 1 mL of Ag(N) NP hydrosol with 10 µL of 16 mM ethanolic solution of TP and 200 µL of ethanol, which was used as an internal standard. The concentration of ethanol was chosen so that the most intense band of ethanol (882 cm$^{-1}$) was the only band that distinguishable in the spectrum. The excitation profile was constructed using a slightly modified version of a previously published protocol [92]: The integral intensities of the spectral bands of TP were normalized with respect to 882 cm$^{-1}$. No further corrections were made.

## 3 Results and Discussion

### 3.1 MPBA Deboronation

Initially, focused on describing how MPBA deboronation varies as a function of experimental conditions (excitation wavelength and morphology of the SERS active system). For this purpose, we prepared the Ag(N)-MPBA system, as described in the experimental section. This system was measured without stirring in a 1.0 cm cuvette at 785 nm laser excitation wavelength ($\lambda_{ex}$) and 8 mW laser power (light red spectrum in **Figure 1**). Immediately after the end of spectrum accumulation, the laser line was switched to 532 nm $\lambda_{ex}$ and 11 mW laser power (approximately 2 min delay) to acquire further spectrum (light green spectrum in **Figure 1**). Immediately afterwards, the sample was measured again at 785 nm (dark red spectrum, **Figure 1**) and then at 532 nm $\lambda_{ex}$ (dark green spectrum) (**Figure 1**).

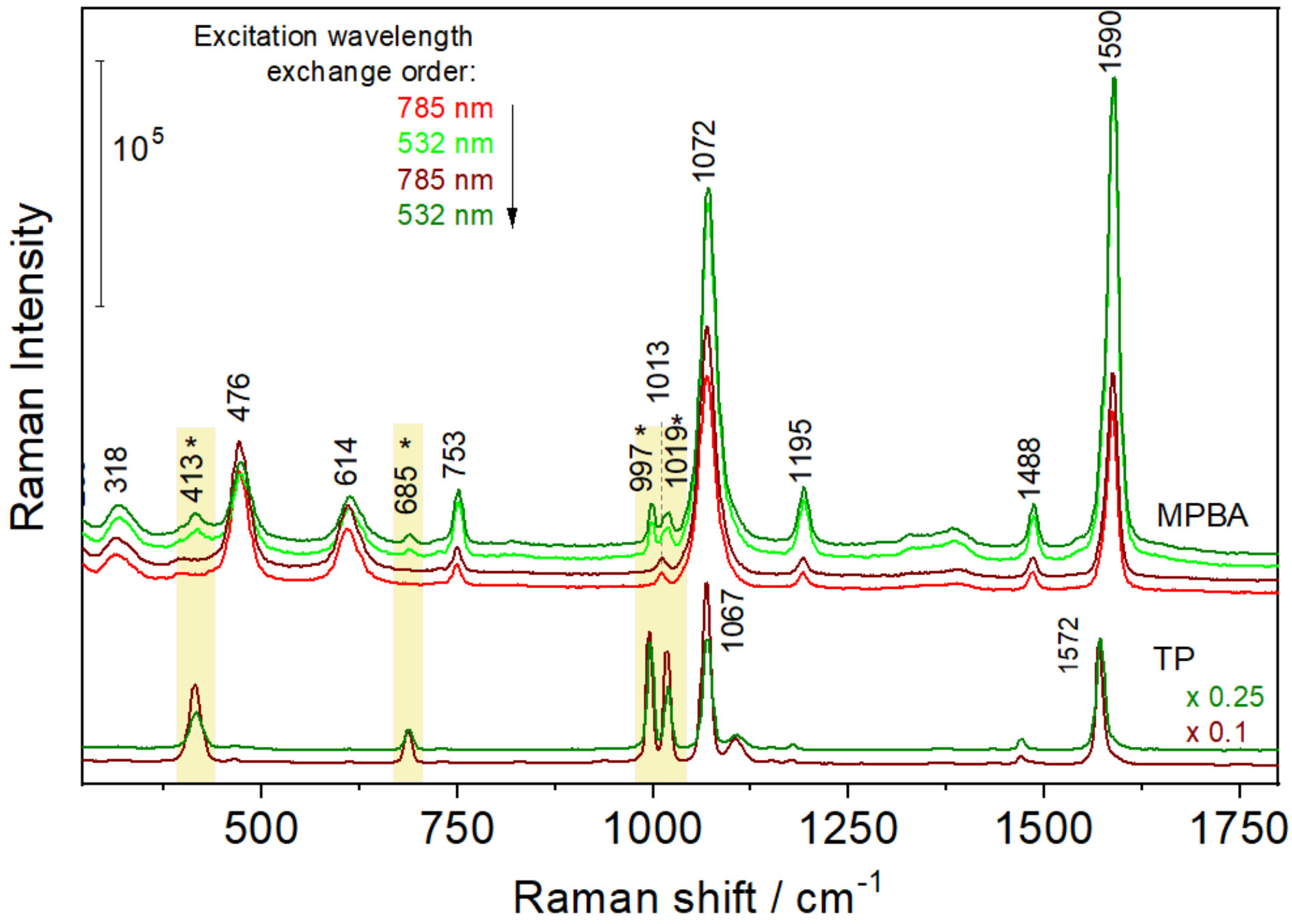

**Figure 1** - Upper SERS spectra show Ag(N)-MPBA system in 1.0 cm cuvette measured alternately with 785 and 532 nm excitation. Lower SERS spectra display Ag(N)-TP system measured with 532 nm (green) and 785 nm (red) excitations

As shown in **Figure 1**, at 532 nm $\lambda_{ex}$, new bands, marked with an asterisk, appeared at - 413, 685, 997 and 1019 cm$^{-1}$. These are characteristic bands of thiophenol (TP), whose SERS spectra (Ag(N)-TP system) measured at both 532 and 785 nm $\lambda_{ex}$ are shown for comparison at the bottom of **Figure 1**. In particular, the doublet 997 and 1019 cm$^{-1}$, which is typical of Raman spectra of monosubstituted benzene derivatives, can be used as evidence of MPBA deboronation. After switching from 532 nm back to 785 nm $\lambda_{ex}$, these bands disappeared again. The appearance of the bands of TP solely at 532 nm cannot be explained by the lower electromagnetic enhancement of the SERS signal [7,14] at 785 nm or by the formation of a surface complex of TP on Ag NPs, leading to an additional chemical enhancement at 532 nm excitation. **Figure S3** shows that the LSPR band of the Ag(N)-MPBA system was found at 752 nm and that the LSPR maximum of Ag(N)-TP system was located at 872 nm, so SERS enhancement of TP should be higher at 785 than 532 nm $\lambda_{ex}$. Any additional chemical enhancement was ruled out for TP by constructing the SERS excitation profile of the Ag(N)-TP system (**Figure S4**). The relative enhancement of all bands matched the LSPR progression, bottoming at 532 nm and peaking at 780 nm. Therefore, the disappearance of the TP signal after switching the laser must have a different explanation.

## 3.2 The experimental setup affects the formation of TP and BPDT from MPBA

Based on the results described above, we hypothesized that deboronation depends on the experimental setup, that is, the deboronation product was formed only at specific $\lambda_{ex}$. We used a standard macro chamber, a 1.0 cm cuvette, and an accumulation time of several seconds to ensure that only a small portion of the aggregates contained in the cuvette was excited. If deboronation occurred at 532 but not at 785 nm $\lambda_{ex}$, the same aggregates would unlikely be irradiated with the laser after switching $\lambda_{ex}$ from 532 to 785 nm. To test this hypothesis, we performed the following experiments in two extreme configurations, in terms of convection and concentration of the laser beam in the cuvette.

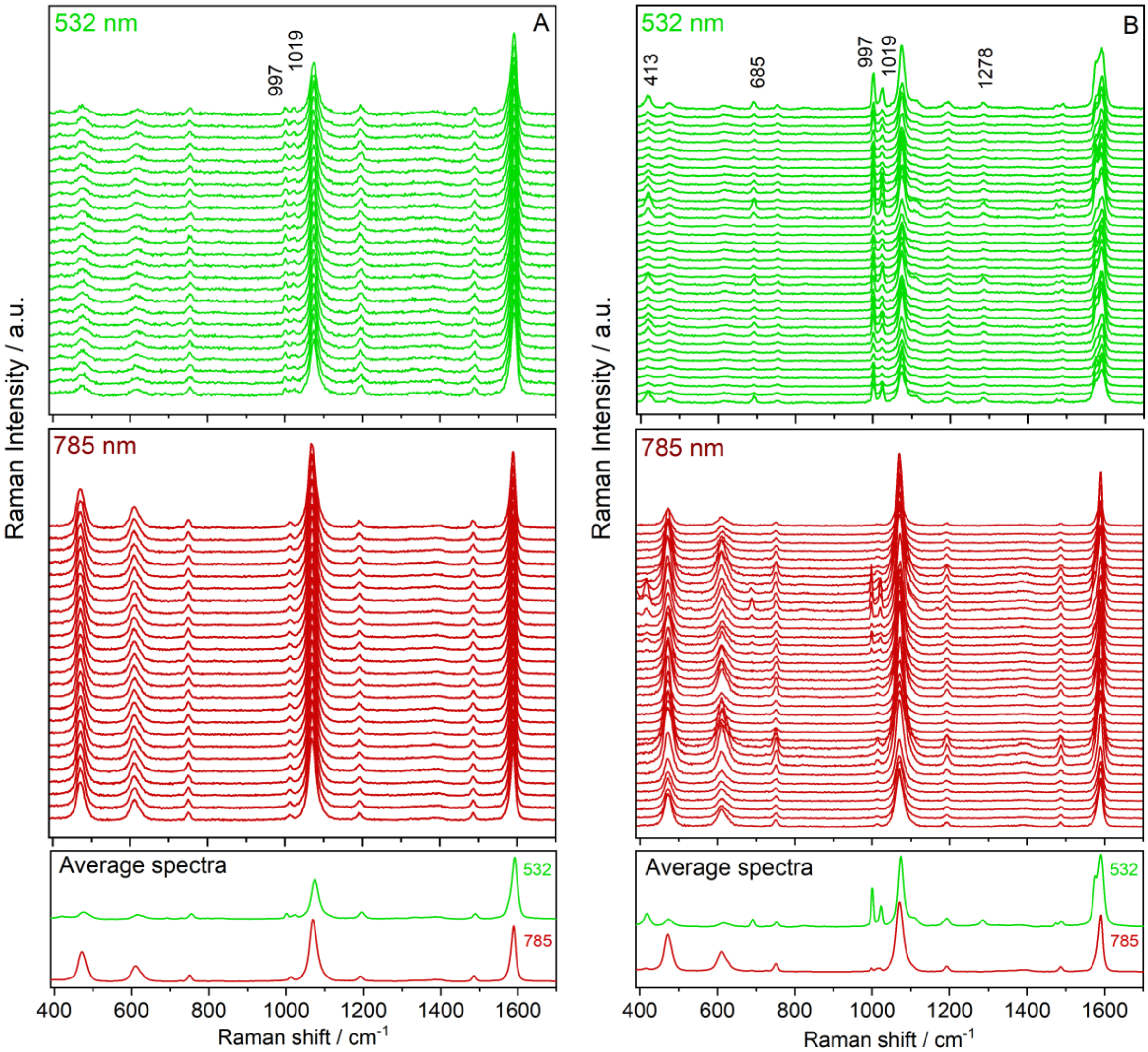

**Figure 2** - Time series of SERS spectra of the Ag(N)-MPBA system A) in a 1.0 cm cuvette with 5 s accumulation time and B) in the confocal setup in 0.2 cm cuvette with 0.5 s accumulation time, and laser excitation wavelengths of 532 and 785 nm, respectively

Spectra of the Ag(N)-MPBA system were measured at both 532 and 785 nm $\lambda_{ex}$: i) for 250 s, in a 1.0 cm cuvette, in a macro chamber, and with an accumulation time of 5 s (**Figure 2A**), and ii) for 100 s, in a 0.2 cm, cuvette with an accumulation time of 0.5 s, and with a 10x objective under a confocal microscope (**Figure 2B**). The spectral series were then processed by factor analysis. No spectral changes (only bands belonging to MPBA) were observed in the spectra of the Ag(N)-MPBA system in a 1.0 cm cuvette with a 5 s accumulation time at 785 nm $\lambda_{ex}$ (**Figure 2A**). FA also provided only a single relevant subspectrum (**Figure S5**) matching the SERS spectrum of MPBA. Although no spectral changes were identified in the spectral set measured at 532 nm excitation (**Figure 2A**) either and the FA of this set (**Figure S6**) also provided a single subspectrum, both the spectra and subspectrum $S_1$ contained traces of deboronation - characteristic bands of TP - 997 and 1019 cm$^{-1}$. **Figure 2B** shows the measurement results in a confocal setup: a thin cuvette with short accumulation times. The spectra at both $\lambda_{ex}$ were not homogeneous, and significant spectral changes occurred in the system **over** time. The new bands detected (at 413, 685 and doublet 997 and 1019 cm$^{-1}$) at 785 nm $\lambda_{ex}$ belonged to TP. These bands appeared and disappeared over time (**Figure 2B**). The FA of this spectral set (**Figure S7**) provided up to 3 relevant subspectra. The first subspectrum, $S_1$, was a weighted average of the spectral intensities and contained both MPBA and TP bands. The coefficients $V_{i1}$ yielded information about the intensity variation of the SERS signal in the time course. Subspectrum $S_2$ contained positive bands belonging to TP and negative bands belonging to MPBA, indicating MPBA deboronation. The coefficients $V_{i2}$ were non-zero only for spectra in which the TP signal was detected. The third subspectrum, $S_3$, contained negative bands in the lower wavenumber region and positive bands in the higher wavenumber region and corresponded to changes in LSPR of the aggregate from which the signal was measured at a given moment [93]. Under the same experimental conditions, the TP signal was higher at 532 than at 785 nm $\lambda_{ex}$ and remained observable throughout the measurement (**Figure 2B**). Its spectral contribution varied with time, as shown by the FA results (**Figure S8**, $V_{i2}$ coefficients). In addition, at this excitation, a new band was observed at 1278 cm$^{-1}$. This band corresponded to C-C bonding between aromatic rings, i.e., to BPDT generation by MPBA homocoupling (**Scheme 2**). The coupling reaction that leads to the formation BPDT on plasmonic nanoparticles has already been reported for aryl halide compounds [58,94] but has been often overlooked in studies on MPBA [73,74]**.** The spectral ratio of TP to MPBA SERS signal increased slightly over time (**Figure 3A**), whereas the ratio between the SERS signal of TP and BPDT remained virtually constant (**Figure 3B**).

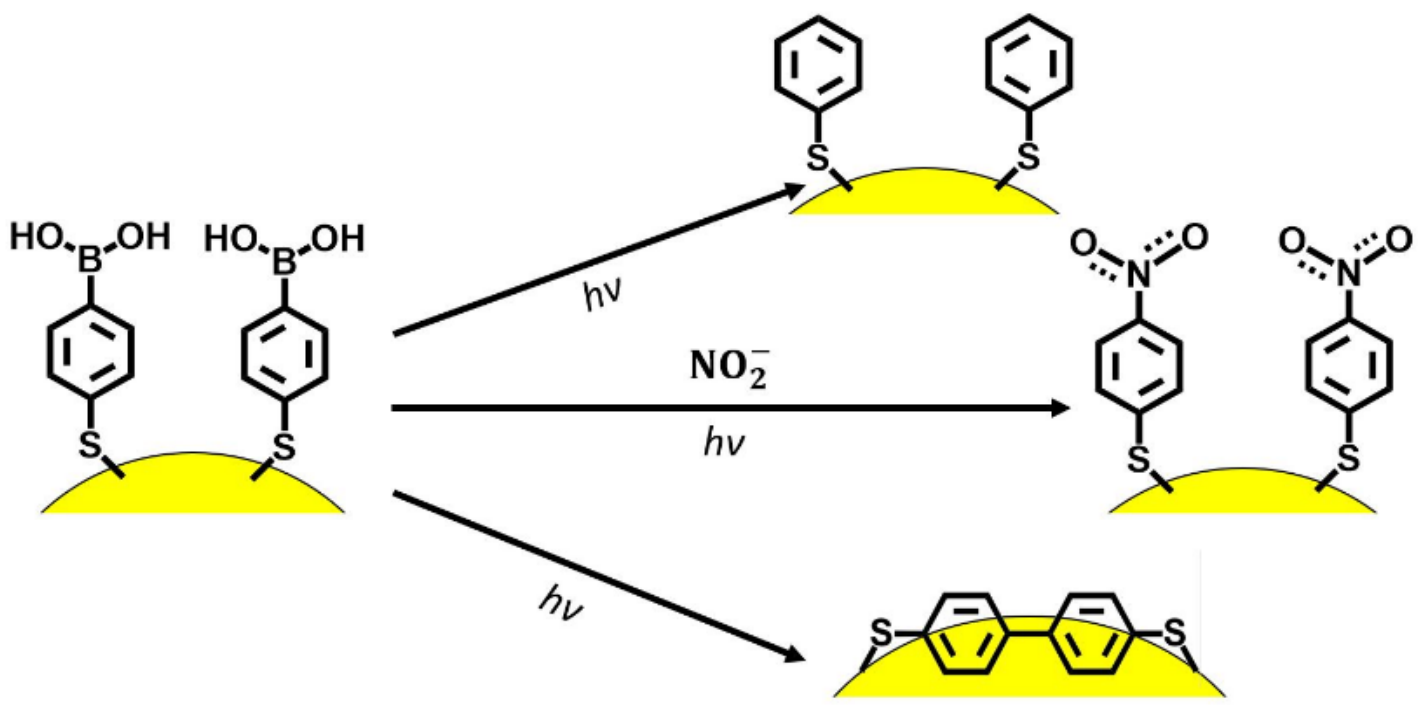

**Scheme 2** – Observed reactions of MPBA on Ag NPs upon illumination with lasers

Based on our findings, we conclude that the degree of MPBA deboronation in Ag NPs hydrosol systems depends not only on $\lambda_{ex}$ but also on the experimental setup, i.e., on the focusing of the laser beam in the cuvette and on the motion of the aggregates in solution. At 532 nm excitation, the deboronation occurred in all experimental arrangements. Whereas at 785 nm excitation, the deboronation was observed in solution only in the confocal setup since the laser beam focus is restricted to a smaller volume compared to the classical "macro" measurement setup. Additionally, in a thin cuvette (0.2 cm thick) the circulation of the sample is lower, which enables to achieve higher laser power per time per an aggregate. Due to the fluctuation of the TP signal at both 532 and 785 nm $\lambda_{ex}$, we suggest that MPBA deboronation was mainly restricted to *hot spots* of aggregates with "suitable" LSPRs illuminated during signal accumulation. For this reason, we measured sample SERS spectra of MPBA, TP, BPDT and NTP (**Figure 4**) at 785 nm excitation in a 1.0 cm cuvette.

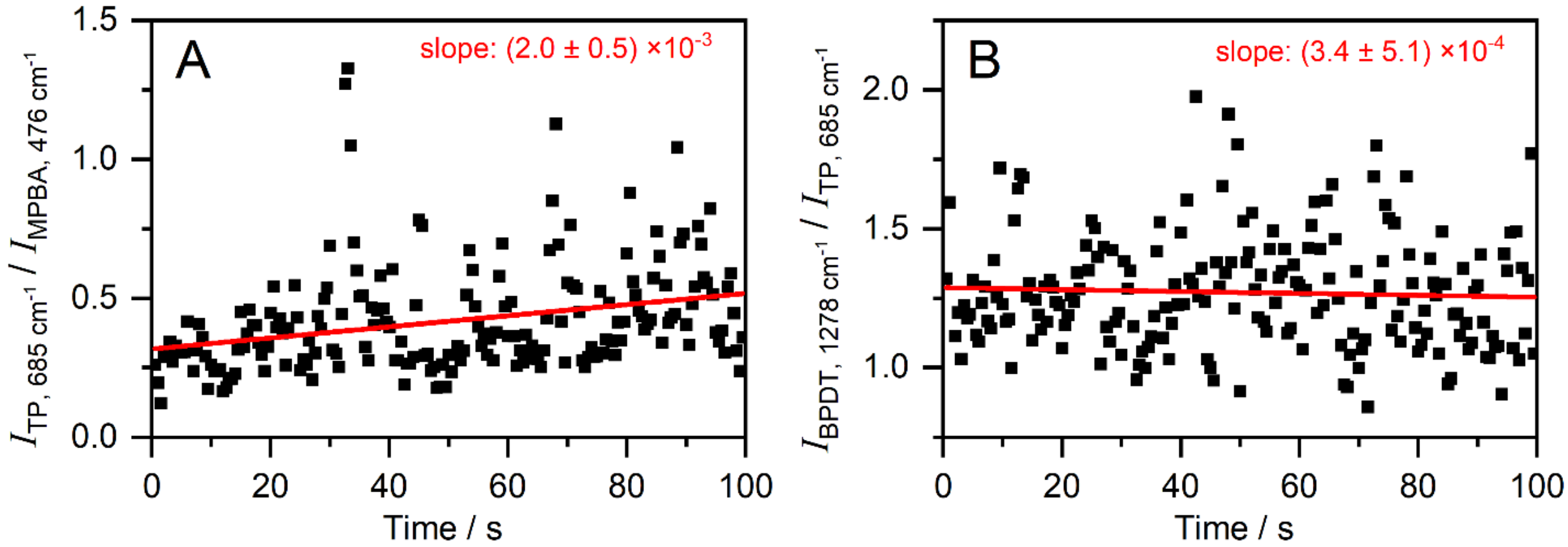

**Figure 3** - The spectral ratio of A) TP vs. MPBA, B) BFDT vs. TP in the time spectral set measured in a 0.2 cm cuvette, 0.5 s accumulation and at 532 nm laser excitation wavelength

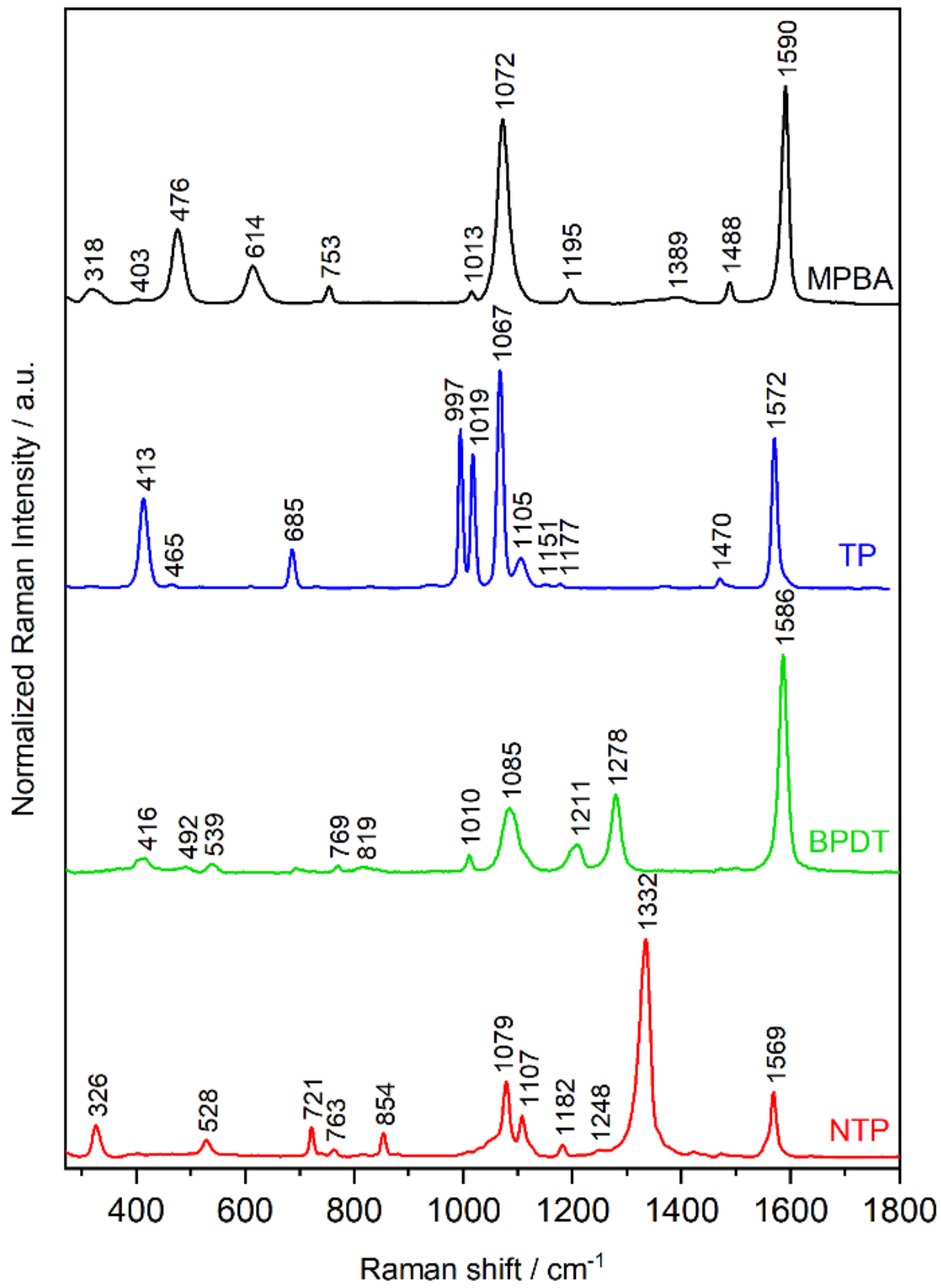

**Figure 4** - SERS spectra of Ag(N)-MPBA, Ag(N)-TP, Ag(N)-BFDT, Ag(N)-NTP systems in a 1.0 cm cuvette and at 785 nm laser excitation wavelength

## 3.3 NTP formation

### 3.3.1 MPBA deboronation upon 445 nm excitation

To analyze in depth the role of $\lambda_{ex}$ in MPBA deboronation, we switched the laser to 445 nm. **Figure 5** shows the result of the SERS spectra time evolution of Ag(N)-MPBA system in a 1.0 cm cuvette and with a 10 s accumulation time. Even the first spectrum contained not only signs of MPBA deboronation, showing the TP bands at 413, 685, 997, 1019 cm$^{-1}$ (marked with an asterisk) and the BPDT band at 1278 cm$^{-1}$ (marked with two asterisks), but also bands with a noticeable progression over time – 326, 721, 854, and 1332 and a shoulder at 1571 cm$^{-1}$. None of these bands has been previously described in SERS spectra of MPBA or its deboronation products.

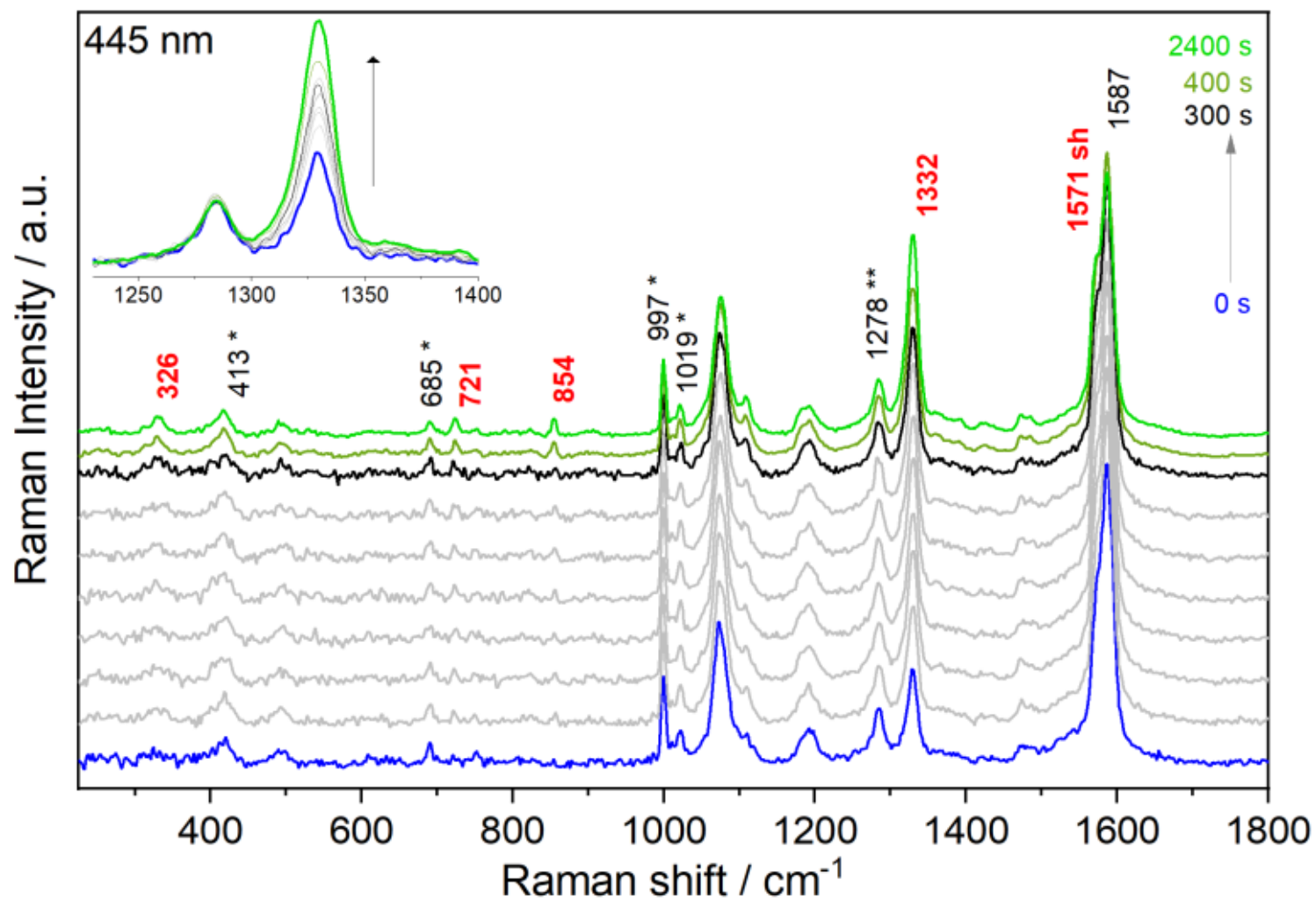

**Figure 5** - SERS spectra of Ag(N)-MPBA system excited at 445 nm. Blue to black – 10 s accumulation, 10 mW, every second spectrum omitted. Green – the same system after 400 and 2400 s, respectively (64 s accumulations). Bands of TP are marked by *, and the band of BPDT is marked by double asterisks

After ruling out any sample contamination, we performed FA on this spectral series. FA (**Figure S9**) provided two relevant subspectra. The 1$^{st}$ subspectrum $S_1$ and its relevant coefficients $V_{i1}$ represented an average spectrum and fluctuation of the SERS signal. The positive bands in subspectrum $S_2$ corresponded to the unknown bands. The variation of $V_{i2}$ coefficients as a function of time confirmed the increase of the SERS signal of this unknown substance.

By linear combination of subspectra $S_1$ and $S_2$ (procedure described in SI), we obtained spectra of the supposedly pure components, $P_1$ and $P_2$. Spectrum $P_1$ consisted of a mix of bands of three substances – MPBA, TP (bands marked *), and BPDT (bands marked **) because the ratio of these substances practically did not change during the measurement. In contrast, the $P_2$ spectrum contained all the previously identified new bands at 326, 721, 854 and 1332 cm$^{-1}$ (**Figure 5**) and other new bands at 1079, 1107 and 1569 cm$^{-1}$. We compared the spectrum $P_2$ with SERS spectra of aromatic benzenethiol derivatives and found a match with the SERS spectrum of NTP shown in **Figure 4** and at the bottom of **Figure 6** (the most intensive band 1332 cm$^{-1}$ belongs to the stretching vibration of $NO_2$ group (ν($NO_2$)).

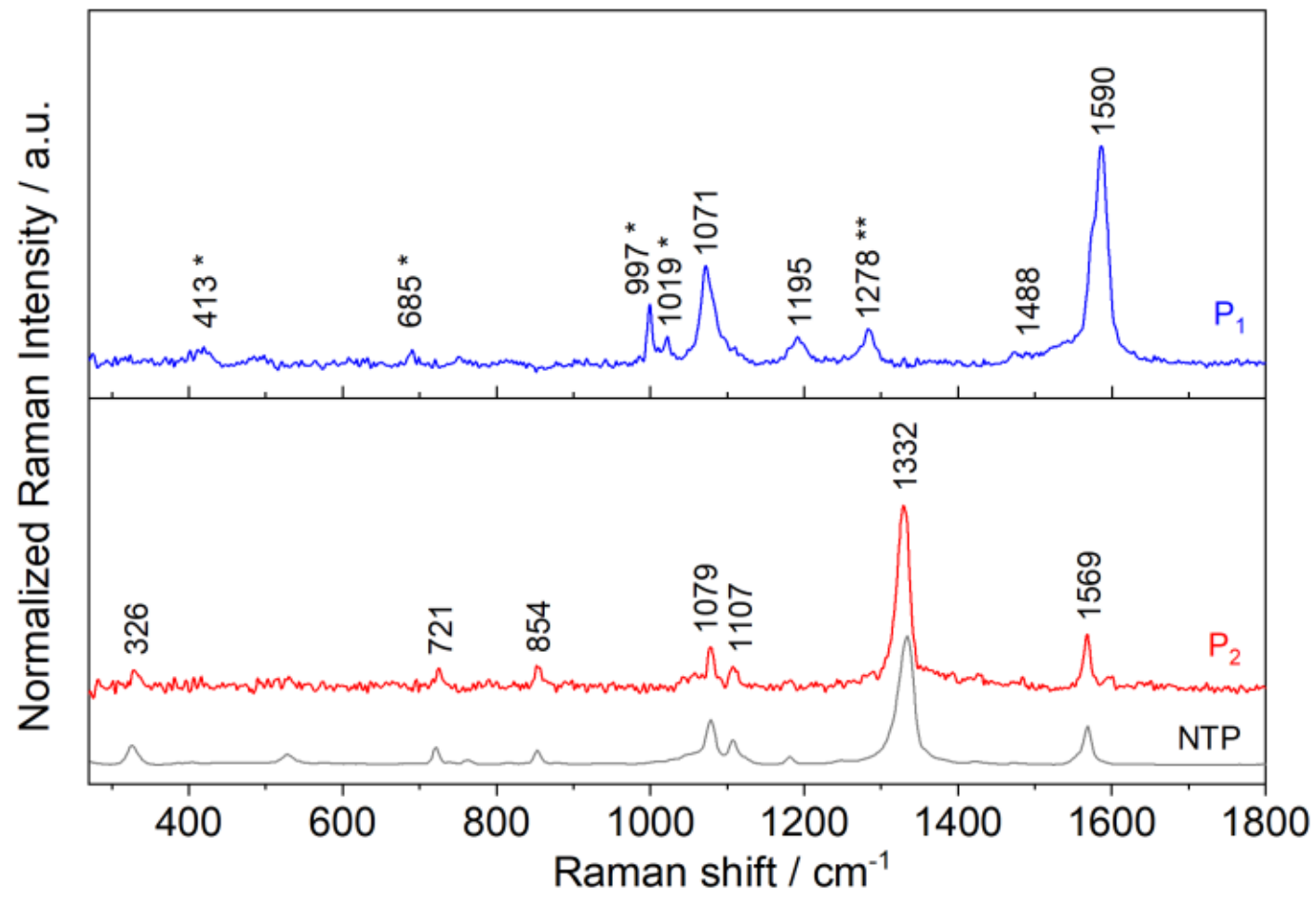

**Figure 6** - Pure components identified by combining the first two subspectra of factor analysis of the time evolution spectra of Ag(N)-MPBA system measured at 445 nm excitation with the SERS spectrum of Ag(N)-NTP system

### 3.3.2 SERS active systems with deposited NPs

In the previous section, we showed that the shorter the wavelength, the greater the beam focus, and the smaller the sample convection are, the greater MPBA deboronation will be. Accordingly, deboronation should peak when using deposited nanoparticle systems. For this reason, we prepared deposited arrays of Ag(N)-MPBA (**Figure S10**).

Although the sample was burning at 445 nm $\lambda_{ex}$ even at the minimum laser power (0.1mW), we were able to record SERS spectra at 532 nm $\lambda_{ex}$ excitation and 0.1 mW (**Figure S11**). From the first spectrum, the characteristic bands of TP, BPDT together with bands of NTP (327 and 1330 cm$^{-1}$) were visible, intensifying over time. At higher laser powers the sample burned. When using 785 nm $\lambda_{ex}$ and the lowest laser power of 0.1 mW, no changes in the SERS spectra of MPBA were observed (**Figure S12**). From 0.3 to the maximum laser power of 1.3 mW, when no sample burning was yet occurring, the TP, BPDT, and 1330 cm$^{-1}$ bands gradually appeared. Thus, the unknown bands that belong to NTP appear in Ag(N)-MPBA systems even when using wavelengths longer than 445 nm on solid substrates.

### 3.3.3 Where does the $NO_2$ group come from?

Considering that the possible source of the $NO_2$ group could be a residue from the silver nitrate used to prepare the hydrosol, we prepared Ag NPs hydrosols in which $AgNO_3$ was reduced by sodium borohydride (Ag(B)) and sodium citrate (Ag(C)) together with an Au NPs reduced by sodium citrate (Au(C)). Because the highest deboronation and BPDT and NTP formation were achieved in the deposited NP systems, we used this sample morphology again, preparing 2D arrays deposited on glass from all these hydrosols and MPBA as an adsorbate.

**Figure 7** shows SERS spectra of the corresponding Ag NPs-MPBA arrays at 532 nm $\lambda_{ex}$ and of the Au(C)-MPBA array at 785 nm $\lambda_{ex}$. For the Au(C)-MPBA array, a longer wavelength was chosen due to the LSPR position of Au NPs. The 1330 cm$^{-1}$ band was not observed in any spectrum of these systems. In Ag NPs – MPBA films, only TP and BPDT bands appeared. In Au(C) – MPBA film, no BPDT formation was observed; only deboronation took place, with the characteristic doublet of TP. Thus, we ruled out $AgNO_3$ as the source of the $NO_2$ group.

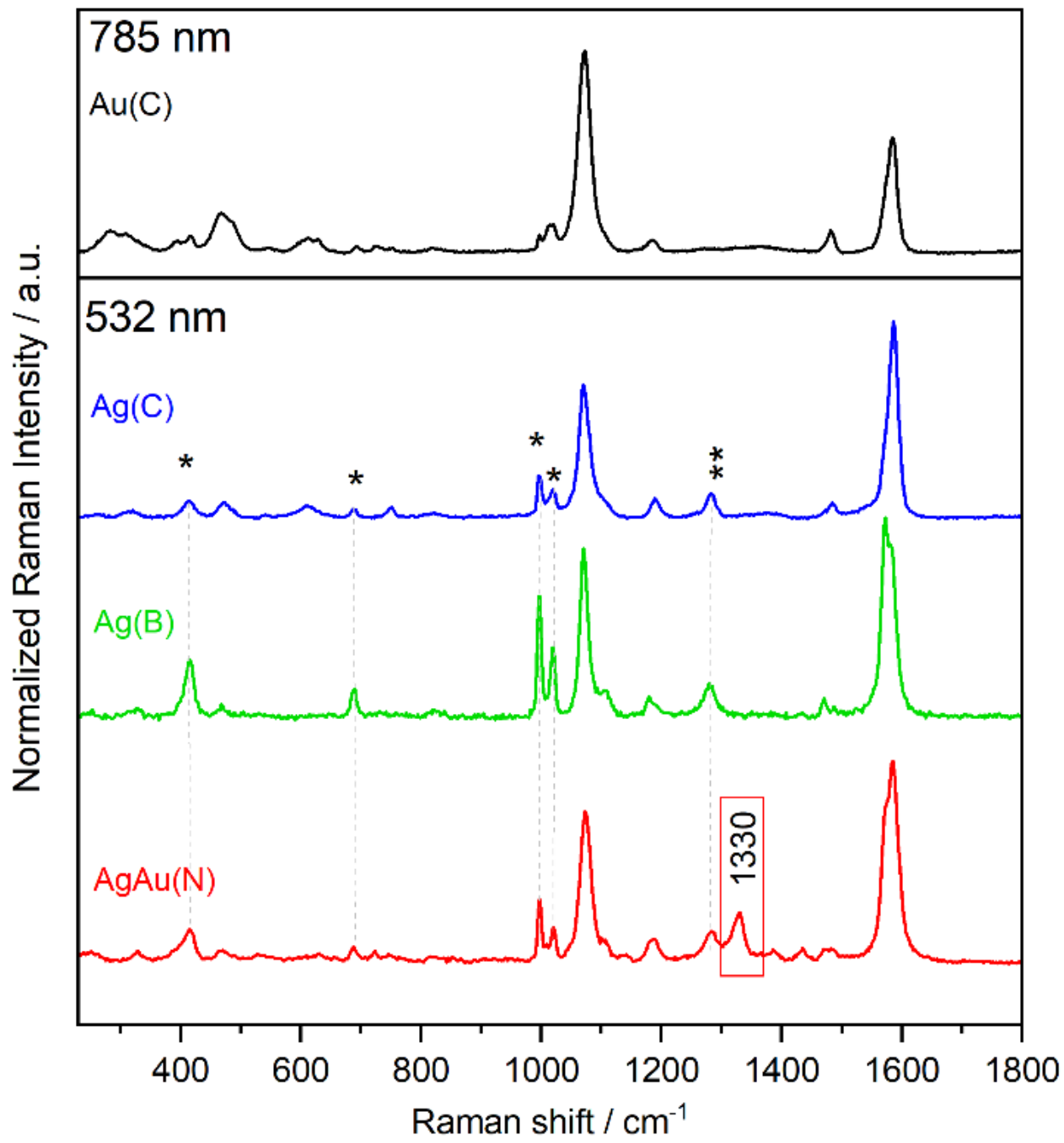

**Figure 7** - SERS spectra of 2D arrays of Au(C)-MPBA (excitation 785 nm) and Ag(C)-MPBA, Ag(B)-MPBA, AgAu-MPBA (all excitation 532 nm).

The only other nitrogen-containing compound used to prepare the Ag(N)-MPBA system was the reducing agent hydroxylamine hydrochloride. According to the literature, [95,96] its autooxidation products are the following (**Scheme 3**):

$$NH_2OH + O_2 \rightarrow NH_2O^{\cdot} + O_2^{\cdot-} + H^+$$
$$NH_2O^{\cdot} + O_2^{\cdot-} \rightarrow HNO + HO_2^-$$
$$HNO + HO_2^- \rightarrow NO^- + H_2O_2$$
$$NO^- + O_2 \rightleftharpoons ONOO^-$$
$$ONOO^- + NH_2OH \rightleftharpoons NO_2^- + NO^- + H_2O + H^+$$
$$2NO^- + 2H^+ \rightleftharpoons N_2O + H_2O$$

**Scheme 3** – Autooxidation of hydroxylamine [95,96]

To confirm that a hydroxylamine oxidation product was our nitrating agent, we tested another NP hydrosol prepared using hydroxylamine hydrochloride ($NH_2OH.HCl$) as a reducing agent. First, we tested the Au(N) NP hydrosol, albeit without aggregation after adding MPBA. This effect was most probably due to high stability of this sol caused by the presence of chloride anions originating from both $HAuCl_4$ and $NH_2OH.HCl$ used in the NPs synthesis. Therefore, we failed to measure the SERS spectra. Subsequently, we tested the bimetallic AgAu(N) NPs hydrosol. The resulting SERS spectrum at 532 nm $\lambda_{ex}$ of this AgAu(N)-MPBA 2D array is shown in **Figure 7**.

Unlike in the other systems shown in **Figure 7**, in this spectrum, the ν($NO_2$) band at 1330 $cm^{-1}$ was present. Thus, the source of the $NO_2$ group was indeed the hydroxylamine. This conclusion was further supported by the fact that the ν($NO_2$) band appeared only for approximately 14 days after preparing the Ag(N) hydrosols. Later, the hydroxylamine in solution seemed to be completely decomposed into the final product $N_2O$ (**Scheme 3**) and 1330 $cm^{-1}$ band was not observed after the illumination by 532 nm.

### 3.3.4 NTP formation from MPBA by $NaNO_2$

In previous experiments, we excluded silver nitrate used in the preparation of hydrosols as a source of $NO_2$ group. Even the extra addition of $NO_3$ ions in the form of $KNO_3$ did not lead to the conversion of MPBA to NTP. It was shown that the source of the $NO_2$ group in the transformation of MPBA to NTP was hydroxylamine hydrochloride, more precisely, one of its oxidation products. Since one of the oxidation products of hydroxylamine (**Scheme 3**) was the nitrite anion, we tested the direct addition of nitrite ions ($NaNO_2$) to the Ag(N)-MPBA system in a 1 cm cuvette.

To this end, we conducted the tests at 532 nm $\lambda_{ex}$ because the previous experiments (**Figure 2A**) showed that no NTP was spontaneously generated in the Ag(N)-MPBA system in this experimental setup. Moreover, due to the resonant Raman effect, NTP traces are easier to detect at this wavelength than at 785 nm. **Figure 8** shows that both deboronation (bands of TP are marked with asterisks) and, to a small extent, BPDT generation (band marked with double asterisks) occurred when adding $NaNO_2$ to the system. However, NTP bands at 1107 $cm^{-1}$, 1569 $cm^{-1}$ and particularly ν($NO_2$) at 1330 $cm^{-1}$ also appeared, and their intensity, unlike that of TP and BPDT bands, depended on the amount of $NaNO_2$ added to the system.

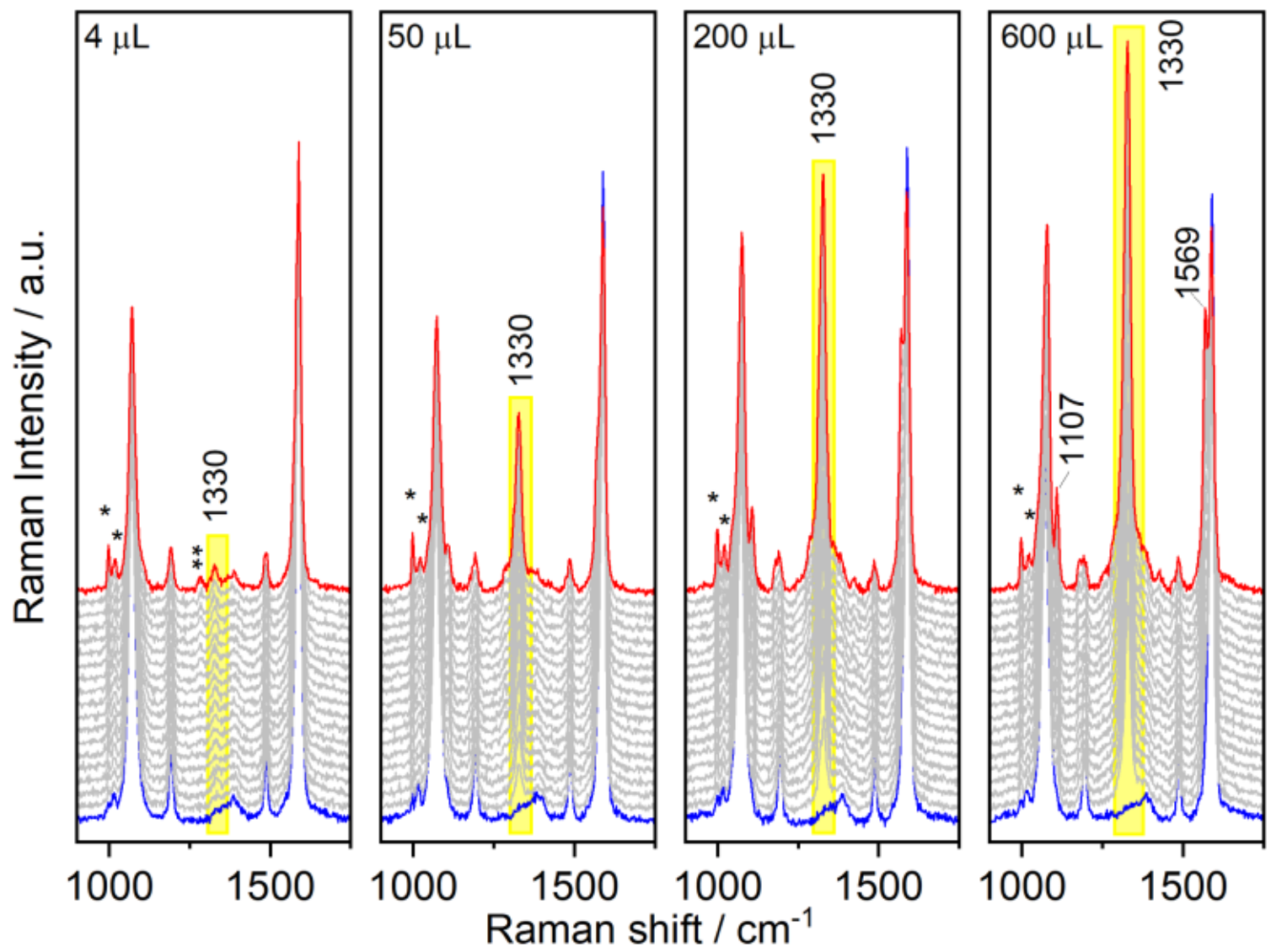

**Figure 8** – The time evolution of the Ag(N)-MPBA system with the addition of $NaNO_2$ solution (the amount of 1 M aqueous solution of $NaNO_2$ is depicted in the upper part of each spectrum), in a 1.0 cm cuvette, at 532 nm laser excitation wavelength, 10 mW, and 10 s accumulation time. Every 5th spectrum is plotted. Bands of TP are marked by an asterisk, and bands of BPDT by double asterisks).

In further experiments, we used Ag(N)-MPBA 2D arrays deposited on a glass slide, performing all measurements at 785 nm $\lambda_{ex}$ and 0.15 mW laser power. Under these conditions, only bands attributable to MPBA were observed in the SERS spectra, with no spectral changes over time (**Figure 9A**). Subsequently, we overlaid the Ag(N)-MPBA 2D array with a $2\times10^{-1}$ M $NaNO_2$ solution and recorded again the time-dependent spectra. **Figure 9B** shows bands belonging to NTP, including the peak ν($NO_2$) at 1332 $cm^{-1}$, which increased in intensity over time, in addition to a weak TP signal (doublet 997 and 1019 $cm^{-1}$). Thus, we were able to demonstrate that the nitrating agent responsible for MPBA conversion into NTP is the nitrite anion.

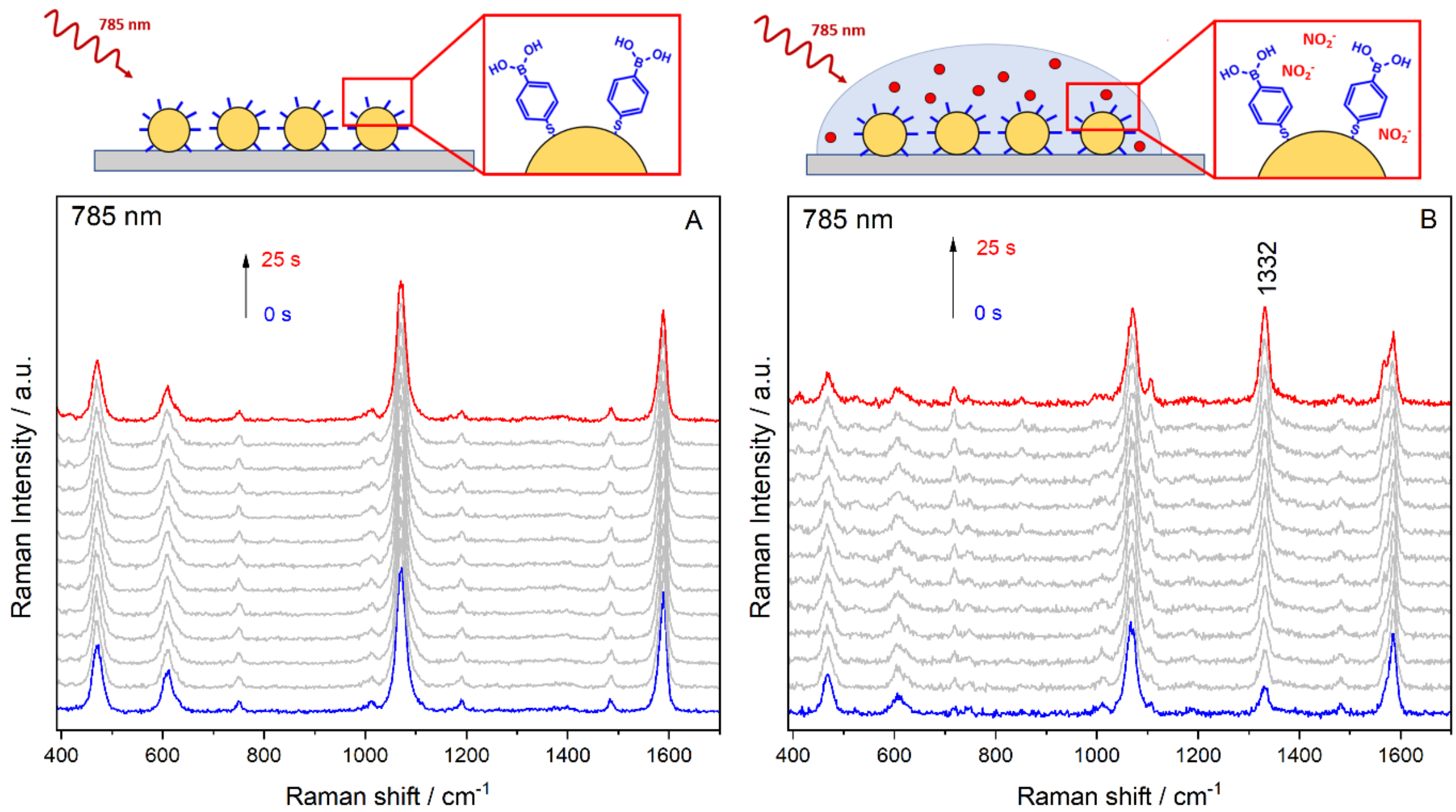

**Figure 9 -** Time evolution of SERS spectra of 2D array Ag(N)-MPBA, at 785 nm laser excitation wavelength, 0.15 mW, accumulation time 0.5 s. A) as prepared, B) covered with a 0.2 M solution of $NaNO_2$ (plus the scheme of the experiment).

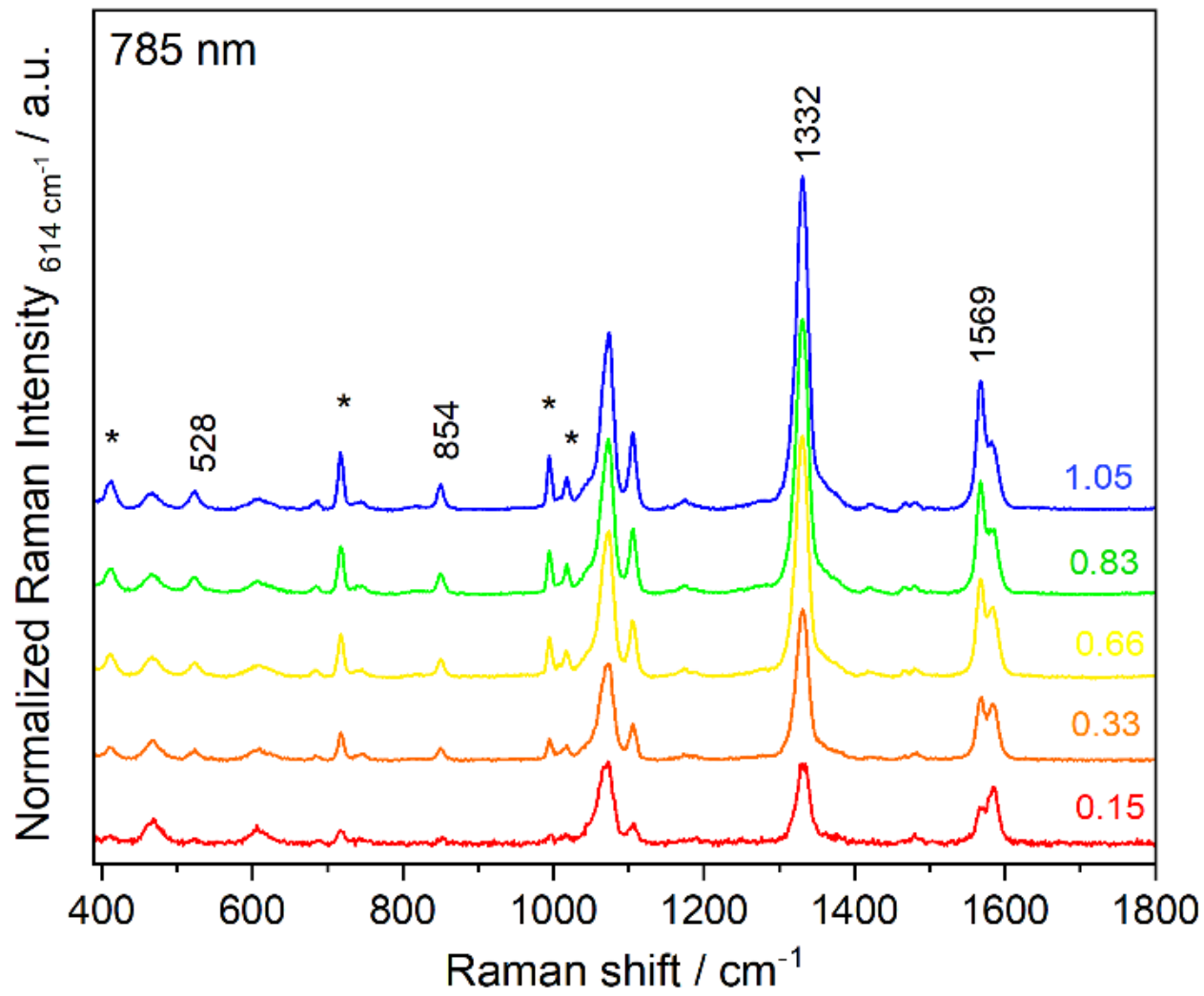

**Figure 10** - SERS spectra of the 2D array Ag(N)-MPBA covered with a 0.2 M solution of $NaNO_2$, at 785 nm laser excitation wavelength, and 0.5 s accumulation time; final spectra after irradiation for 25 s. The laser power is indicated in the image

Although a detailed analysis of the NTP formation mechanism goes beyond the scope of this study, in the last step, we tried to determine the effect of the laser power on the formation of NTP in the system shown in **Figure 9B**. **Figure 10** shows the final SERS spectra obtained after 25 s of signal accumulation with the laser power in the range from 0.15 to 1.05 mW. As shown in this figure, the NTP signal (bands at 528, 854, 1332 and 1569 cm$^{-1}$) in the system increased with laser power relative to the MPBA signal, with a concurrent increase in TP formation (bands marked with an asterisk). The whole time-evolution SERS spectra are shown in **Figure S13A** together with plots of integral intensities of ν($NO_2$) ($I_{1332}$) and MPBA band at 475 cm$^{-1}$ ($I_{475}$) as a function of time (**Figure S12B** and **C**, respectively). Because the $NO_2^-$ concentration was significantly higher than the MPBA concentration, we assumed pseudo first-order kinetics.

Based on these results, both curves of the variation of $I_{1332}$ and $I_{475}$ as a function of time were fitted with an exponential function, thereby deriving the rate constants ($k_{NTP}$ and $k_{MPBA}$) (**Figure S13**). The values of the rate constants increased with the laser power and ranged from $3.6\times10^{-2}$ $s^{-1}$ to $1.02\times10^{-1}$ $s^{-1}$ for $k_{NTP}$ and 3. $\times10^{-2}$ $s^{-1}$ to $1.38\times10^{-1}$ $s^{-1}$ for $k_{MPBA}$. In **Figure 11**, both $k_{NTP}$ and $k_{MPBA}$ are plotted as a function of laser power. The MPBA extinction rate constants $k_{MPBA}$ were higher than the NTP formation rate constant $k_{NTP}$ at all laser powers. Simultaneously, as the laser power increased, so did their difference. This finding is consistent with the fact that other parallel deboronation reactions (TP and BPDT formation) occurred in addition to NTP formation. Combined, these results indicate that laser excitation is required for MPBA conversion into NTP on the surface of Ag NPs.

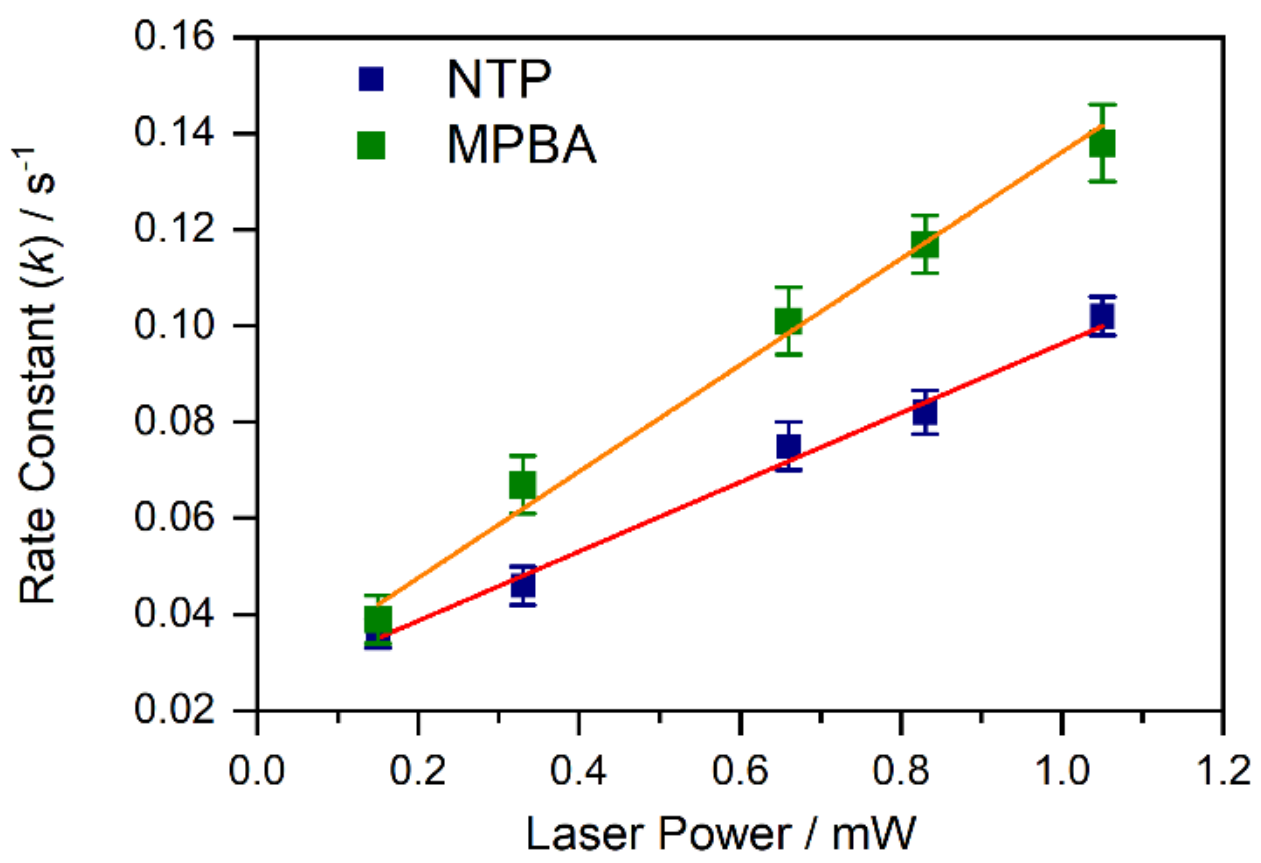

**Figure 11** - Rate constants of NTP formation in 2D array Ag(N)-MPBA covered with a 0.2 M solution of $NaNO_2$ as a function of laser power. Error bars were calculated from 4 repeated experiments.

The reaction rates of plasmon-catalyzed reactions vary with the laser power. This relationship differs for reactions controlled by the photothermal effect and for reactions controlled by hot carriers [18,19,97,98]. For reactions mediated by plasmonic heating, the reaction rate should be an exponential function of the laser power [18,97]. For reactions controlled by hot carriers, a linear [20] or super-linear [19,99] (high laser power) function should express the relationship between the reaction rate and the laser power . Since the most reasonable goodness-of-fit of the curves of NTP formation rate (given by $k_{NTP}$) and MPBA decomposition rate (given by $k_{MPBA}$) as a function of laser power shown in **Figure 11** was a straight line, we suppose that MPBA conversion into NTP due to the reaction with $NO_2^-$ ions is driven by hot carrier effects.

## 4 Conclusions

The experimental setup is crucial for plasmon-driven reactions of MPBA adsorbed on the surface of Ag NPs. In liquid systems at 785 nm excitation, no or only small-scale deboronation occurs, and MBPA remains in its "native" state. However, when using a shorter wavelength or a solid AgNP-MPBA system, deboronation occurs, and both TP and BPDT are detected in SERS spectra. Suppressing the deboronation reaction requires using a minimal laser power (lower than 0.2 mW at the sample) and 785 nm $\lambda_{ex}$. These results have considerable practical implications because many applications involving the detection of biologically important molecules are based on MPBA interactions linked to Ag NPs deposited systems with 1,2- or 1,3-diols via the -$B(OH)_2$ group. Thus, if deboronation does not derive from interactions with diols, but from laser irradiation, the results may be highly biased. Moreover, after irradiating liquid Ag(N) NPs-MPBA at a 445 nm or solid Ag(N) NPs-MPBA systems at longer $\lambda_{ex}$, a new photochemical product of MPBA appears in SERS spectra. The positions of its SERS bands match those of 4-nitrothiophenol (NTP) in these spectra. NTP is a reaction product of the interaction of MPBA with an oxidation product of $NH_2OH.HCl$ used as the reduction agent during the preparation of Ag NPs. MPBA conversion into NTP on the surface of Ag NPs can also be mediated by sodium nitrite addition. The increased intensity of NTP bands over time correlates with the decrease in MPBA signal. The rate constants of NTP formation as a function of laser power follows a linear trend, suggesting the prevalence of hot carrier effects in this plasmon-driven reaction. The actual mechanism of this reaction will be the subject of further research since the mechanisms of -$B(OH)_2$ replacement by -$NO_2$ in organic chemistry are based on either radical or electrophilic substitution involving $NO_2^+$ [82]. In this event, NTP results from the reaction of the $NO_2^-$ anion with MPBA adsorbed on Ag NPs.

***CRediT authorship contribution statement***

**JK**: Investigation, Visualization, Resources, Writing-Reviewing and Editing. **JH:** Investigation. **MS**: Investigation, Writing-Reviewing and Editing. **IS:** Supervision of the experiments, Project administration, Conceptualization, Methodology, Investigation, Formal Analysis, Visualization, Writing – original draft.

***Declaration of Competing Interest***

The authors declare that they have no known competing financial interests or personal relationships that could have appeared to influence the work reported in this paper.

## Acknowledgements

This work was supported by the Charles University Grant Agency (GAUK) 237823 and the authors acknowledge the assistance provided by the Advanced Multiscale Materials for Key Enabling Technologies project, supported by the Ministry of Education, Youth, and Sports of the Czech Republic. Project No. CZ.02.01.01/00/22_008/0004558, Co-funded by the European Union.
The authors gratefully thank Dr. Carlos V. Melo for editing the manuscript.

## Appendix A. Supplementary material

Supplementary material available in a separate file.

## References

[1] U. Kreibig, M. Vollmer, Optical Properties of Metal Clusters, 1st ed., Springer-Verlag Berlin Heidelberg, 1995. https://doi.org/10.1007/978-3-662-09109-8.

[2] K.L. Kelly, E. Coronado, L.L. Zhao, G.C. Schatz, The optical properties of metal nanoparticles: The influence of size, shape, and dielectric environment, J. Phys. Chem. B. 107 (2003) 668–677. https://doi.org/10.1021/jp026731y.

[3] R.F. Aroca, Plasmon enhanced spectroscopy, Phys. Chem. Chem. Phys. 15 (2013) 5355–5363.

https://doi.org/10.1039/c3cp44103b.

[4] M. Moskovits, B.D. Piorek, A brief history of surface-enhanced Raman spectroscopy and the localized surface plasmon Dedicated to the memory of Richard Van Duyne (1945–2019), J. Raman Spectrosc. 52 (2021) 279–284. https://doi.org/10.1002/jrs.6028.

[5] X. Wang, S.C. Huang, S. Hu, S. Yan, B. Ren, Fundamental understanding and applications of plasmon-enhanced Raman spectroscopy, Nat. Rev. Phys. 2 (2020) 253–271. https://doi.org/10.1038/s42254-020-0171-y.

[6] H. Yu, Y. Peng, Y. Yang, Z.Y. Li, Plasmon-enhanced light–matter interactions and applications, Npj Comput. Mater. 5 (2019) 1–14. https://doi.org/10.1038/s41524-019-0184-1.

[7] S.Y. Ding, E.M. You, Z.Q. Tian, M. Moskovits, Electromagnetic theories of surface-enhanced Raman spectroscopy, Chem. Soc. Rev. 46 (2017) 4042–4076. https://doi.org/10.1039/c7cs00238f.

[8] E.C. Le Ru, P.G. Etchegoin, Single-molecule surface-enhanced raman spectroscopy, Annu. Rev. Phys. Chem. 63 (2012) 65–87. https://doi.org/10.1146/annurev-physchem-032511-143757.

[9] H. Xu, M. Käll, Polarization-dependent surface-enhanced raman spectroscopy of isolated silver nanoaggregates, ChemPhysChem. 4 (2003) 1001–1005. https://doi.org/10.1002/cphc.200200544.

[10] M. Moskovits, Surface-enhanced Raman spectroscopy: a brief retrospective, J. Raman Spectrosc. 36 (2005) 485–496. https://doi.org/10.1002/jrs.1362.

[11] J. Langer, D.J. de Aberasturi, J. Aizpurua, R.A. Alvarez-Puebla, B. Auguié, J.J. Baumberg, G.C. Bazan, S.E.J. Bell, A. Boisen, A.G. Brolo, J. Choo, D. Cialla-May, V. Deckert, L. Fabris, K. Faulds, F. Javier García de Abajo, R. Goodacre, D. Graham, A.J. Haes, C.L. Haynes, C. Huck, T. Itoh, M. Käll, J. Kneipp, N.A. Kotov, H. Kuang, E.C. Le Ru, H.K. Lee, J.F. Li, X.Y. Ling, S.A. Maier, T. Mayerhöfer, M. Moskovits, K. Murakoshi, J.M. Nam, S. Nie, Y. Ozaki, I. Pastoriza-Santos, J. Perez-Juste, J. Popp, A. Pucci, S. Reich, B. Ren, G.C. Schatz, T. Shegai, S. Schlücker, L.L. Tay, K. George Thomas, Z.Q. Tian, R.P. van Duyne, T. Vo-Dinh, Y. Wang, K.A. Willets, C. Xu, H. Xu, Y. Xu, Y.S. Yamamoto, B. Zhao, L.M. Liz-Marzán, Present and future of surface-enhanced Raman scattering, ACS Nano. 14 (2020) 28–117. https://doi.org/10.1021/acsnano.9b04224.

[12] H. Kneipp, Katrin; Moskovits, Maritn; Kneipp, ed., Surface-enhanced Raman scattering: Physics and applications, 1st ed., Springer-Verlag Berlin Heidelberg, 2006. https://doi.org/10.1007/3-540-33567-6.

[13] D. Cialla, A. Maerz, R. Boehme, F. Theil, K. Weber, M. Schmitt, J. Popp, Surface-enhanced Raman spectroscopy (SERS): progress and trends, Anal. Bioanal. Chem. 403 (2012) 27–54. https://doi.org/10.1007/s00216-011-5631-x.

[14] R. Aroca, Surface-Enhanced Vibrational Spectroscopy, John Wiley and Sons, Chichester, UK, 2006. https://doi.org/10.1002/9780470035641.

[15] J.P. Scaffidi, M.K. Gregas, V. Seewaldt, T. Vo-Dinh, SERS-based plasmonic nanobiosensing in single living cells, Anal. Bioanal. Chem. 393 (2009) 1135–1141. https://doi.org/10.1007/s00216-008-2521-y.

[16] A. Manjavacas, J.G. Liu, V. Kulkarni, P. Nordlander, Plasmon-Induced Hot Carriers in Metallic Nanoparticles, ACS Nano. 8 (2014) 7630–7638. https://doi.org/10.1021/nn502445f.

[17] J. Zhao, J. Wang, A.J. Brock, H. Zhu, Plasmonic heterogeneous catalysis for organic transformations, J. Photochem. Photobiol. C Photochem. Rev. 52 (2022) 100539.

https://doi.org/10.1016/j.jphotochemrev.2022.100539.

[18] M.J. Kale, T. Avanesian, P. Christopher, Direct Photocatalysis by Plasmonic Nanostructures, ACS Catal. 4 (2014) 116–128. https://doi.org/10.1021/cs400993w.

[19] S. Linic, U. Aslam, C. Boerigter, M. Morabito, Photochemical transformations on plasmonic metal nanoparticles, Nat. Mater. 14 (2015) 567–576. https://doi.org/10.1038/NMAT4281.

[20] S. Mukherjee, F. Libisch, N. Large, O. Neumann, L. V Brown, J. Cheng, B. Lassiter, E.A. Carter, P. Nordlander, N.J. Halas, Hot Electrons Do the Impossible: Plasmon-Induced Dissociation of $H_2$ on Au, Nano Lett. 13 (2013) 240–247. https://doi.org/10.1021/nl303940z.

[21] Z. Li, D. Kurouski, Nanoscale Structural Characterization of Plasmon-Driven Reactions, Nanophotonics. 10 (2021) 1657–1673. https://doi.org/10.1515/nanoph-2020-0647.

[22] G. Baffou, I. Bordacchini, A. Baldi, R. Quidant, Simple experimental procedures to distinguish photothermal from hot-carrier processes in plasmonics, Light Sci. Appl. 9 (2020) 1–16. https://doi.org/10.1038/s41377-020-00345-0.

[23] Y. Dubi, I.W. Un, Y. Sivan, Thermal effects - an alternative mechanism for plasmon-assisted photocatalysis, Chem. Sci. 11 (2020) 5017–5027. https://doi.org/10.1039/c9sc06480j.

[24] C. Zhan, J. Yi, S. Hu, X.-G. Zhang, D.-Y. Wu, Z.-Q. Tian, Plasmon-mediated chemical reactions, Nat. Rev. METHODS Prim. 3 (2023). https://doi.org/10.1038/s43586-023-00195-1.

[25] M.J. Landry, A. Gellé, B.Y. Meng, C.J. Barrett, A. Moores, Surface-Plasmon-Mediated Hydrogenation of Carbonyls Catalyzed by Silver Nanocubes under Visible Light, ACS Catal. 7 (2017) 6128–6133. https://doi.org/10.1021/acscatal.7b02128.

[26] Q. Xiao, S. Sarina, E.R. Waclawik, J. Jia, J. Chang, J.D. Riches, H. Wu, Z. Zheng, H. Zhu, Alloying Gold with Copper Makes for a Highly Selective Visible-Light Photocatalyst for the Reduction of Nitroaromatics to Anilines, ACS Catal. 6 (2016) 1744–1753. https://doi.org/10.1021/acscatal.5b02643.

[27] H. Zhu, X. Ke, X. Yang, S. Sarina, H. Liu, Reduction of nitroaromatic compounds on supported gold nanoparticles by visible and ultraviolet light, Angew. Chemie - Int. Ed. 49 (2010) 9657–9661. https://doi.org/10.1002/anie.201003908.

[28] A. Stefancu, J. Gargiulo, G. Laufersky, B. Auguié, V. Chiş, E.C. Le Ru, M. Liu, N. Leopold, E. Cortés, Interface-Dependent Selectivity in Plasmon-Driven Chemical Reactions, ACS Nano. 17 (2023) 3119–3127. https://doi.org/10.1021/acsnano.2c12116.

[29] A.G.M. da Silva, T.S. Rodrigues, V.G. Correia, T. V. Alves, R.S. Alves, R.A. Ando, F.R. Ornellas, J. Wang, L.H. Andrade, P.H.C. Camargo, Plasmonic Nanorattles as Next-Generation Catalysts for Surface Plasmon Resonance-Mediated Oxidations Promoted by Activated Oxygen, Angew. Chemie - Int. Ed. 55 (2016) 7111–7115. https://doi.org/10.1002/anie.201601740.

[30] R.K. Rai, D. Tyagi, K. Gupta, S.K. Singh, Activated nanostructured bimetallic catalysts for C-C coupling reactions: Recent progress, Catal. Sci. Technol. 6 (2016) 3341–3361. https://doi.org/10.1039/c5cy02225h.

[31] R. Bayan, N. Karak, Photo-Assisted Synthesis of a Pd-Ag@CQD Nanohybrid and Its Catalytic Efficiency in Promoting the Suzuki-Miyaura Cross-Coupling Reaction under Ligand-Free and Ambient Conditions, ACS Omega. 2 (2017) 8868–8876. https://doi.org/10.1021/acsomega.7b01504.

[32] O. Guselnikova, P. Postnikov, M.M. Chehimi, Y. Kalachyovaa, V. Svorcik, O. Lyutakov, Surface

Plasmon-Polariton: A Novel Way to Initiate Azide-Alkyne Cycloaddition, Langmuir. 35 (2019) 2023–2032. https://doi.org/10.1021/acs.langmuir.8b03041.

[33] C.F. Wang, B.T. O'Callahan, D. Kurouski, A. Krayev, P.Z. El-Khoury, The Prevalence of Anions at Plasmonic Nanojunctions: A Closer Look at p-Nitrothiophenol, J. Phys. Chem. Lett. 11 (2020) 3809–3814. https://doi.org/10.1021/acs.jpclett.0c01006.

[34] R. Schürmann, T.F.M. Luxford, I.S. Vinklárek, J. Kočišek, M. Zawadzki, I. Bald, Interaction of 4-nitrothiophenol with low energy electrons: Implications for plasmon mediated reactions, J. Chem. Phys. 153 (2020) 104303. https://doi.org/10.1063/5.0018784.

[35] R. Schürmann, K. Ebel, C. Nicolas, A.R. Milosavljević, I. Bald, Role of Valence Band States and Plasmonic Enhancement in Electron-Transfer-Induced Transformation of Nitrothiophenol, J. Phys. Chem. Lett. 10 (2019) 3153–3158. https://doi.org/10.1021/acs.jpclett.9b00848.

[36] Y. Liu, J. Deng, Z. Jin, T. Liu, J. Zhou, F. Luo, G. Wang, C. Xjefmz, B. Jo, D. Qiztjdt, B.O.E. Cjpmphz, A study of plasmon-driven catalytic 4-NBT to DMAB in the dry film by using spatial Raman mapping spectroscopy, Nano Res. 15 (2022) 6062–6066. https://doi.org/10.1007/s12274-022-4310-x.

[37] Y. Liu, L. Zhang, X. Liu, Y. Zhang, Y. Yan, Y. Zhao, In situ SERS monitoring of plasmon-driven catalytic reaction on gap-controlled Ag nanoparticle arrays under 785 nm irradiation, Spectrochim. Acta - Part A Mol. Biomol. Spectrosc. 270 (2022) 120803. https://doi.org/10.1016/j.saa.2021.120803.

[38] Y. Ling, W.C. Xie, G.K. Liu, R.W. Yan, D.Y. Wu, J. Tang, The discovery of the hydrogen bond from p-Nitrothiophenol by Raman spectroscopy: Guideline for the thioalcohol molecule recognition tool, Sci. Rep. 6 (2016) 6–10. https://doi.org/10.1038/srep31981.

[39] Z. Li, C. Zhang, H. Sheng, J. Wang, Y. Zhu, L. Yu, J. Wang, Q. Peng, G. Lu, Molecular Cocatalyst of p-Mercaptophenylboronic Acid Boosts the Plasmon-Mediated Reduction of p-Nitrothiophenol, ACS Appl. Mater. Interfaces. 14 (2022) 38302–38310. https://doi.org/10.1021/acsami.2c08327.

[40] K. Kim, K.L. Kim, K.S. Shin, Photoreduction of 4,4′-dimercaptoazobenzene on Ag revealed by Raman scattering spectroscopy, Langmuir. 29 (2013) 183–190. https://doi.org/10.1021/la304159c.

[41] J.L. Brooks, R.R. Frontiera, Competition between Reaction and Degradation Pathways in Plasmon-Driven Photochemistry, J. Phys. Chem. C. 120 (2016) 20869–20876. https://doi.org/10.1021/acs.jpcc.6b02314.

[42] Y. Zhu, H. Tang, H. Wang, Y. Li, *In Situ* SERS Monitoring of the Plasmon-Driven Catalytic Reaction by Using Single Ag@Au Nanowires as Substrates, Anal. Chem. 93 (2021) 11736–11744. https://doi.org/10.1021/acs.analchem.1c01926.

[43] M. Yang, Y. Pan, C. Ji, M. Shao, Z. Li, J. Yu, C. Li, B. Man, C. Zhang, X. Zhao, Three-Dimensional MXene-AgNP Hollow Spheres for In Situ Surface-Enhanced Raman Scattering Detection of Catalysis Reactions, J. Phys. Chem. Lett. 14 (2023) 9019–9026. https://doi.org/10.1021/acs.jpclett.3c02379.

[44] W. Xie, S. Schlücker, Hot electron-induced reduction of small molecules on photorecycling metal surfaces, Nat. Commun. 6 (2015) 1–6. https://doi.org/10.1038/ncomms8570.

[45] T. Touzalin, S. Joiret, E. Maisonhaute, I.T. Lucas, Complex Electron Transfer Pathway at a Microelectrode Captured by in Situ Nanospectroscopy, Anal. Chem. 89 (2017) 8974–8980.

https://doi.org/10.1021/acs.analchem.7b01542.

[46] D.S. Lopes, L.L.R. Vono, E. V. Miranda, R.A. Ando, P. Corio, Inhibition of p-Nitrothiophenol Catalytic Hydrogenation on Ag-Containing AgAu/Pd/TiO2 Plasmonic Catalysts Probed in situ by SERS, ChemCatChem. 14 (2022) e2021019. https://doi.org/10.1002/cctc.202101943.

[47] H. Huh, H.D. Trinh, D. Lee, S. Yoon, How Does a Plasmon-Induced Hot Charge Carrier Break a C-C Bond?, ACS Appl. Mater. Interfaces. 11 (2019) 24715–24724. https://doi.org/10.1021/acsami.9b05509.

[48] L.T.M. Huynh, H.D. Trinh, S. Lee, S. Yoon, Plasmon-driven protodeboronation reactions in nanogaps, Nanoscale. 12 (2020) 24062–24069. https://doi.org/10.1039/d0nr07023h.

[49] G. Li, R. Jin, Catalysis by gold nanoparticles: Carbon-carbon coupling reactions, Nanotechnol. Rev. 2 (2013) 529–545. https://doi.org/10.1515/ntrev-2013-0020.

[50] K. Lacina, P. Skládal, T.D. James, Boronic acids for sensing and other applications - a mini-review of papers published in 2013, Chem. Cent. J. 8 (2014) 1–17. https://doi.org/10.1186/s13065-014-0060-5.

[51] R. Nishiyabu, Y. Kubo, T.D. James, J.S. Fossey, Boronic acid building blocks: Tools for sensing and separation, Chem. Commun. 47 (2011) 1106–1123. https://doi.org/10.1039/c0cc02920c.

[52] F. Sun, D.D. Galvan, P. Jain, Q. Yu, Multi-functional, thiophenol-based surface chemistry for surface-enhanced Raman spectroscopy, Chem. Commun. 53 (2017) 4550–4561. https://doi.org/10.1039/c7cc01577a.

[53] J.C. Love, L.A. Estroff, J.K. Kriebel, R.G. Nuzzo, G.M. Whitesides, Self-assembled monolayers of thiolates on metals as a form of nanotechnology, Chem. Rev. 105 (2005) 1103–1169. https://doi.org/10.1021/cr0300789.

[54] N. de J. Hiller, N.A. do Amaral e Silva, T.A. Tavares, R.X. Faria, M.N. Eberlin, D. de Luna Martins, Arylboronic Acids and their Myriad of Applications Beyond Organic Synthesis, European J. Org. Chem. 2020 (2020) 4841–4877. https://doi.org/10.1002/ejoc.202000396.

[55] D. Sun, G. Qi, S. Xu, W. Xu, Construction of highly sensitive surface-enhanced Raman scattering (SERS) nanosensor aimed for the testing of glucose in urine, RSC Adv. 6 (2016) 53800–53803. https://doi.org/10.1039/c6ra06223g.

[56] N.H. Ly, A.T.N. Lam, N. Dinh Bao, Y.J. Kwark, S.W. Joo, Glucose-induced and fructose-induced deboronation reaction of 4-mercaptophenylboronic acid assembled on silver investigated by surface-enhanced Raman scattering, Surf. Interface Anal. 49 (2017) 495–502. https://doi.org/10.1002/sia.6184.

[57] A. Michota, J. Bukowska, Surface-enhanced Raman scattering (SERS) of 4-mercaptobenzoic acid on silver and gold substrates, J. Raman Spectrosc. 34 (2003) 21–25. https://doi.org/10.1002/jrs.928.

[58] P. Jiang, Y. Dong, L. Yang, Y. Zhao, W. Xie, Hot Electron-Induced Carbon-Halogen Bond Cleavage Monitored by in Situ Surface-Enhanced Raman Spectroscopy, J. Phys. Chem. C. 123 (2019) 16741–16746. https://doi.org/10.1021/acs.jpcc.9b03238.

[59] R. Schürmann, A. Dutta, K. Ebel, K. Tapio, A.R. Milosavljević, I. Bald, Plasmonic reactivity of halogen thiophenols on gold nanoparticles studied by SERS and XPS, J. Chem. Phys. 157 (2022) 084708. https://doi.org/10.1063/5.0098110.

[60] C.G.T. Feugmo, V. Liégeois, Analyzing the vibrational signatures of thiophenol adsorbed on

small gold clusters by DFT calculations, ChemPhysChem. 14 (2013) 1633–1645. https://doi.org/10.1002/cphc.201201077.

[61] R. Holze, The adsorption of thiophenol on gold - A spectroelectrochemical study, Phys. Chem. Chem. Phys. 17 (2015) 21364–21372. https://doi.org/10.1039/c5cp00884k.

[62] S. Li, D. Wu, X. Xu, R. Gu, Theoretical and experimental studies on the adsorption behavior of thiophenol on gold nanoparticles, J. Raman Spectrosc. 38 (2007) 1436–1443. https://doi.org/10.1002/jrs.1791.

[63] S.K. Saikin, R. Olivares-Amaya, D. Rappoport, M. Stopa, A. Aspuru-Guzik, On the chemical bonding effects in the Raman response: Benzenethiol adsorbed on silver clusters, Phys. Chem. Chem. Phys. 11 (2009) 9401–9411. https://doi.org/10.1039/b906885f.

[64] C.A. Szafranski, W. Tanner, P.E. Laibinis, R.L. Garrell, Surface-enhanced Raman spectroscopy of aromatic thiols and disulfides on gold electrodes, Langmuir. 14 (1998) 3570–3579. https://doi.org/10.1021/la9702502.

[65] N. Valley, N. Greeneltch, R.P. Van Duyne, G.C. Schatz, A look at the origin and magnitude of the chemical contribution to the enhancement mechanism of surface-enhanced Raman spectroscopy (SERS): Theory and experiment, J. Phys. Chem. Lett. 4 (2013) 2599–2604. https://doi.org/10.1021/jz4012383.

[66] E.D. Emmons, J.A. Guicheteau, A.W. Fountain, A. Tripathi, Effect of substituents on surface equilibria of thiophenols and isoquinolines on gold substrates studied using surface-enhanced Raman spectroscopy, Phys. Chem. Chem. Phys. 22 (2020) 15953–15965. https://doi.org/10.1039/d0cp01125h.

[67] X. Jiang, Z. Tan, L. Lin, J. He, C. He, B.D. Thackray, Y. Zhang, J. Ye, Surface-Enhanced Raman Nanoprobes with Embedded Standards for Quantitative Cholesterol Detection, Small Methods. 2 (2018) 1800182. https://doi.org/10.1002/smtd.201800182.

[68] N.L. Nechaeva, I.A. Boginskaya, A. Ivanov V, A.K. Sarychev, A. Eremenko V, I.A. Ryzhikov, A.N. Lagarkov, I.N. Kurochkin, Multiscale flaked silver SERS-substrate for glycated human albumin biosensing, Anal. Chim. Acta. 1100 (2020) 250–257. https://doi.org/10.1016/j.aca.2019.11.072.

[69] R. Peng, Y. Si, T. Deng, J. Zheng, J. Li, R. Yang, W. Tan, A novel SERS nanoprobe for the ratiometric imaging of hydrogen peroxide in living cells, Chem. Commun. 52 (2016) 8553–8556. https://doi.org/10.1039/c6cc03412h.

[70] F. Sun, T. Bai, L. Zhang, J.-R. Ella-Menye, S. Liu, A.K. Nowinski, S. Jiang, Q. Yu, Sensitive and Fast Detection of Fructose in Complex Media via Symmetry Breaking and Signal Amplification Using Surface-Enhanced Raman Spectroscopy, Anal. Chem. 86 (2014) 2387–2394. https://doi.org/10.1021/ac4040983.

[71] Y. Zhao, L. Du, H. Li, W. Xie, J. Chen, Is the Suzuki-Miyaura Cross-Coupling Reaction in the Presence of Pd Nanoparticles Heterogeneously or Homogeneously Catalyzed? An Interfacial Surface-Enhanced Raman Spectroscopy Study, J. Phys. Chem. Lett. 10 (2019) 1286–1291. https://doi.org/10.1021/acs.jpclett.9b00351.

[72] S. Zhou, X. Guo, H. Huang, X. Huang, X. Zhou, Z. Zhang, G. Sun, H. Cai, H. Zhou, P. Sun, Triple-Function Au-Ag-Stuffed Nanopancakes for SERS Detection,Discrimination, and Inactivation of Multiple Bacteria, Anal. Chem. 94 (2022) 5785–5796. https://doi.org/10.1021/acs.analchem.1c04920.

[73] H. Gebavi, V. Gasparic, D. Risovic, N. Baran, P.H. Albrycht, M. Ivanda, Features and advantages of flexible silicon nanowires for SERS applications, Beilstein J. Nanotechnol. 10 (2019) 725–734. https://doi.org/10.3762/bjnano.10.72.

[74] W. Guan, G. Chen, Z. Yang, J. Shen, T. Yang, C. Ma, L. Li, H. Gao, Y. Xiong, C. Zhu, Y. Zhou, W. Zhang, Sensitive Determination of Stevioside by Surface-Enhanced Raman Spectroscopy (SERS) Using 4-Mercaptophenylboric Acid Functionalized Silver Nanoparticles (Ag NPs), Anal. Lett. 57 (2023) 742–752. https://doi.org/10.1080/00032719.2023.2222198.

[75] J. Ham, B.J. Yun, W.-G. Koh, SERS-based biosensing platform using shape-coded hydrogel microparticles incorporating silver nanoparticles, Sensors Actuators, B Chem. 341 (2021) 129989. https://doi.org/10.1016/j.snb.2021.129989.

[76] M. Li, D. Lu, R. You, H. Shen, L. Zhu, Q. Lin, Y. Lu, Surface-Enhanced Raman Scattering Biosensor Based on Self-Assembled Gold Nanorod Arrays for Rapid and Sensitive Detection of Tyrosinase, J. Phys. Chem. C. 126 (2022) 12651–12659. https://doi.org/10.1021/acs.jpcc.2c03408.

[77] S. Li, Q. Zhou, W. Chu, W. Zhao, J. Zheng, Surface-enhanced Raman scattering behaviour of 4-mercaptophenyl boronic acid on assembled silver nanoparticles, Phys. Chem. Chem. Phys. 17 (2015) 17638–17645. https://doi.org/10.1039/c5cp02409a.

[78] Y. Lu, T. Zhou, R. You, Y. Wu, H. Shen, S. Feng, J. Su, Fabrication and Characterization of a Highly-Sensitive Surface-Enhanced Raman Scattering Nanosensor for Detecting Glucose in Urine, NANOMATERIALS. 8 (2018) 629. https://doi.org/10.3390/nano8080629.

[79] G.Q. Wallace, M. Tabatabaei, M.S. Zuin, M.S. Workentin, F. Lagugné-Labarthet, A nanoaggregate-on-mirror platform for molecular and biomolecular detection by surface-enhanced Raman spectroscopy, Anal. Bioanal. Chem. 408 (2016) 609–618. https://doi.org/10.1007/s00216-015-9142-z.

[80] Y. Xu, M.M. Hassan, A. Zhu, H. Li, Q. Chen, Dual-mode of magnetic assisted Au@Ag SERS tags and cationic conjugated UCNPs for qualitative and quantitative analysis of multiple foodborne pathogens, Sensors Actuators, B Chem. 344 (2021) 130305. https://doi.org/10.1016/j.snb.2021.130305.

[81] J. Ye, Y. Chen, Z. Liu, A Boronate Affinity Sandwich Assay: An Appealing Alternative to Immunoassays for the Determination of Glycoproteins, Angew. Chemie - Int. Ed. 53 (2014) 10386–10389. https://doi.org/10.1002/anie.201405525.

[82] J.I. Murray, M. V. Silva Elipe, K.D. Baucom, D.B. Brown, K. Quasdorf, S. Caille, Ipso Nitration of Aryl Boronic Acids Using Fuming Nitric Acid, J. Org. Chem. 87 (2022) 1977–1985. https://doi.org/10.1021/acs.joc.1c00886.

[83] S. Manna, S. Maity, S. Rana, S. Agasti, D. Maiti, ipso-Nitration of Arylboronic Acids with Bismuth Nitrate and Perdisulfate, Org. Lett. 14 (2012) 1736–1739. https://doi.org/10.1021/ol300325t.

[84] G. Yan, M. Yang, Recent advances in the synthesis of aromatic nitro compounds, Org. Biomol. Chem. 11 (2013) 2554–2566. https://doi.org/10.1039/c3ob27354g.

[85] K. Zhang, A. Budinská, A. Passera, D. Katayev, N-Nitroheterocycles: Bench-Stable Organic Reagents for Catalytic Ipso-Nitration of Aryl- And Heteroarylboronic Acids, Org. Lett. 22 (2020) 2714–2719. https://doi.org/10.1021/acs.orglett.0c00671.

[86] I. Sloufova, M. Slouf, B. Vlckova, V. Gajdosova, J. Zednik, J. Vohlidal, Controlled Tuning of the

Size of Ag-Hydrosol Nanoparticles by Nonstabilized THF and Detection of Peroxides in THF, LANGMUIR. 35 (2019) 9831–9840. https://doi.org/10.1021/acs.langmuir.9b01449.

[87] N. Leopold, B. Lendl, A new method for fast preparation of highly surface-enhanced Raman scattering (SERS) active silver colloids at room temperature by reduction of silver nitrate with hydroxylamine hydrochloride, J. Phys. Chem. B. 107 (2003) 5723–5727. https://doi.org/10.1021/jp027460u.

[88] P.C. Lee, D. Meisel, Adsorption and Surface-Enhanced Raman of Dyes on Silver and Gold Sols, J. Phys. Chem. 86 (1982) 3391–3395. https://doi.org/10.1021/j100214a025.

[89] I. Srnova-Sloufova, F. Lednicky, A. Gemperle, J. Gemperlova, Core-shell (Ag)Au bimetallic nanoparticles: Analysis of transmissionelectron microscopy images, LANGMUIR. 16 (2000) 9928–9935. https://doi.org/10.1021/la0009588.

[90] M. Pruskova, V. Sutrova, M. Slouf, B. Vlckova, J. Vohlidal, I. Sloufova, Arrays of Ag and Au Nanoparticles with Terpyridine- and Thiophene-Based Ligands: Morphology and Optical Responses, LANGMUIR. 33 (2017) 4146–4156. https://doi.org/10.1021/acs.langmuir.7b00126.

[91] J. Palacký, P. Mojzeš, J. Bok, SVD-based method for intensity normalization, background correction and solvent subtraction in Raman spectroscopy exploiting the properties of water stretching vibrations, J. Raman Spectrosc. 42 (2011) 1528–1539. https://doi.org/10.1002/jrs.2896.

[92] I. Srnova-Sloufova, B. Vlckova, T.L. Snoeck, D.J. Stufkens, P. Matejka, Surface-enhanced Raman scattering and surface-enhanced resonance Raman scattering excitation profiles of Ag-2,2'-bipyridine surface complexes and of $[Ru(bpy)_3]^{2+}$ on Ag colloidal surfaces: Manifestations of the charge-transfer resonance contribution, Inorg. Chem. 39 (2000) 3551–3559. https://doi.org/10.1021/ic991372f.

[93] J. Kozisek, M. Slouf, I. Sloufova, Factor analysis of the time series of SERS spectra reveals water arrangement and surface plasmon changes in Ag nanoparticle systems, Spectrochim. Acta Part A-Molecular Biomol. Spectrosc. 293 (2023) 122454. https://doi.org/10.1016/j.saa.2023.122454.

[94] Y. Li, Y. Hu, F. Shi, H. Li, W. Xie, J. Chen, C−H Arylation on Nickel Nanoparticles Monitored by In Situ Surface-Enhanced Raman Spectroscopy, Angew. Chemie - Int. Ed. 58 (2019) 9049–9053. https://doi.org/10.1002/anie.201902825.

[95] K.G.G.C. De Silva, M.I. Ranasinghe, S. Chowdhury, Understanding the induction time associated with the photoredution of resazurin by hydroxylamine in the presence of gold nanoparticles as a photocatalyst, React. Kinet. Mech. Catal. 131 (2020) 965–977. https://doi.org/10.1007/s11144-020-01875-9.

[96] M.N. Hughes, H.G. Nicklin, Autoxidation of hydroxylamine in alkaline solutions, J. Chem. Soc. A Inorganic, Phys. Theor. Chem. (1971) 164–168. https://doi.org/10.1039/J19710000164.

[97] S.C. Huang, X. Wang, Q.Q. Zhao, J.F. Zhu, C.W. Li, Y.H. He, S. Hu, M.M. Sartin, S. Yan, B. Ren, Probing nanoscale spatial distribution of plasmonically excited hot carriers, Nat. Commun. 11 (2020) 4211. https://doi.org/10.1038/s41467-020-18016-4.

[98] L. Zhou, D.F. Swearer, C. Zhang, H. Robatjazi, H. Zhao, L. Henderson, L. Dong, P. Christopher, E.A. Carter, P. Nordlander, N.J. Halas, Quantifying hot carrier and thermal contributions in plasmonic photocatalysis, Science. 362 (2018) 69–72. https://doi.org/10.1126/science.aat6967.

[99] P. Christopher, H. Xin, A. Marimuthu, S. Linic, Singular characteristics and unique chemical bond activation mechanisms of photocatalytic reactions on plasmonic nanostructures, Nat. Mater. 11 (2012) 1044–1050. https://doi.org/10.1038/NMAT3454.

Supporting information

# *Plasmon-driven substitution of 4-mercaptophenylboronic acid to 4-nitrothiophenol monitored by surface-enhanced Raman spectroscopy*

Jan Kozisek[1], Jana Hrncirova[1], Miroslav Slouf[2], and Ivana Sloufova[1]*

* E-mail for Ivana Sloufova: sloufovi@natur.cuni.cz

**Text S1 – Baseline correction and factor analysis**

The sets of the time-evolution SERS spectra were baseline-corrected and analyzed using the 'singular value decomposition' algorithm and the factor analysis (FA) procedure. This procedure provides matrices of orthonormal subspectra $S_j(\nu)$, orthonormal coefficients $V_{ij}$ and singular numbers $W_j$. Each particular experimental spectrum $Y_i(\nu)$ of the set can be then approximated as:

$$Y_i(\nu) = \sum_{j=1}^{M} V_{ij} \cdot W_j \cdot S_j(\nu)$$

where M is the factor dimension equal to the number of independent components (sub-spectra) resolvable in the spectral set under analysis. The M value can be derived from the elbow point in the plot of singular values $W_j$ as a function of the sub-spectrum number or from the plot of the residual error value. The first sub-spectrum ($S_1$) is a weighted average of the spectral intensities of all included spectra, while the second sub-spectrum ($S_2$) shows major spectral changes within the spectral set. Subtle changes are involved in further subspectra. When only two relevant subspectra were acquired, spectra of pure components were constructed: a proper linear combination of the subspectra $S_1$ and $S_2$ so that no negative peaks were observed yields spectra of pure components involved in the monitored changes.

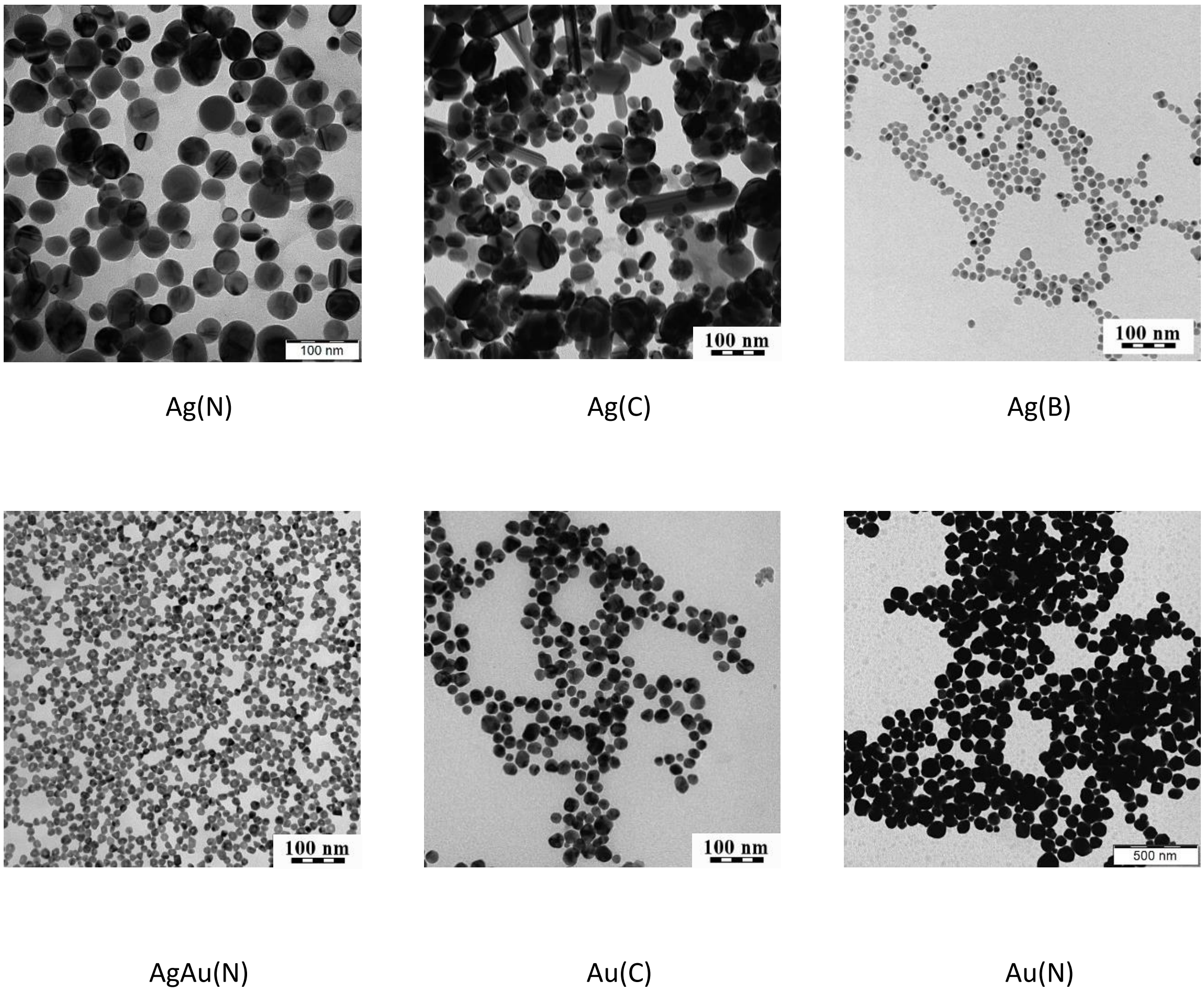

**Figure S1** – Transmission electron microscope (TEM) images of Ag and Au nanoparticles prepared in this study: Ag(N) = reduction agent – hydroxylamine hydrochloride; Ag(C) = reduction agent – trisodium citrate; Ag(B) = reduction agent – sodium borohydride; AgAu(N) = reduction agent – hydroxylamine hydrochloride; Au(C) = reduction agent – trisodium citrate; Au(N) = reduction agent – hydroxylamine hydrochloride

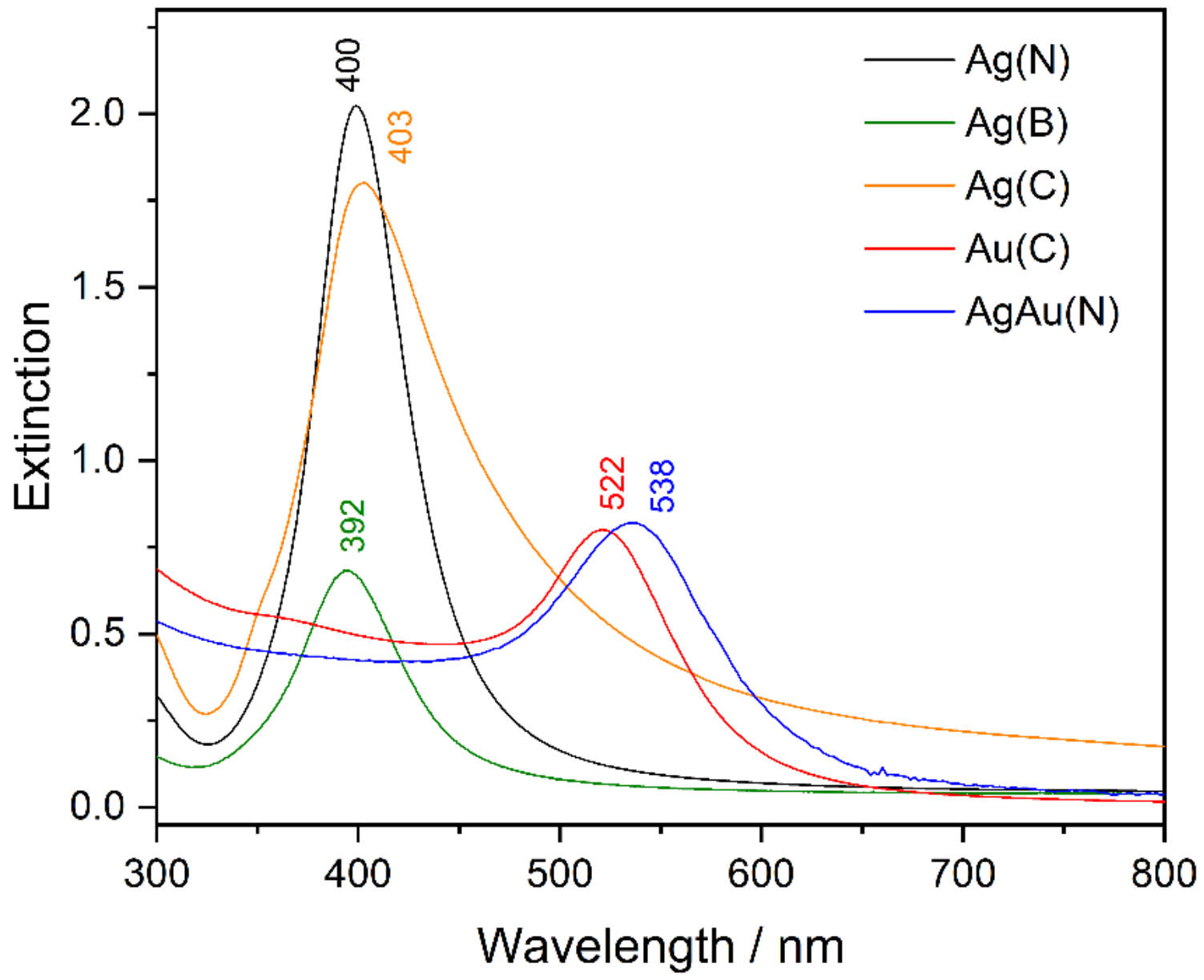

**Figure S2** – Surface plasmon extinction (SPE) spectra of Ag and Au nanoparticles prepared in this study: Ag(N) = reduction agent – hydroxylamine hydrochloride; Ag(C) = reduction agent – trisodium citrate; Ag(B) = reduction agent – sodium borohydride; AgAu(N) = reduction agents – sodium borohydride and hydroxylamine hydrochloride; Au(C) = reduction agent – trisodium citrate.

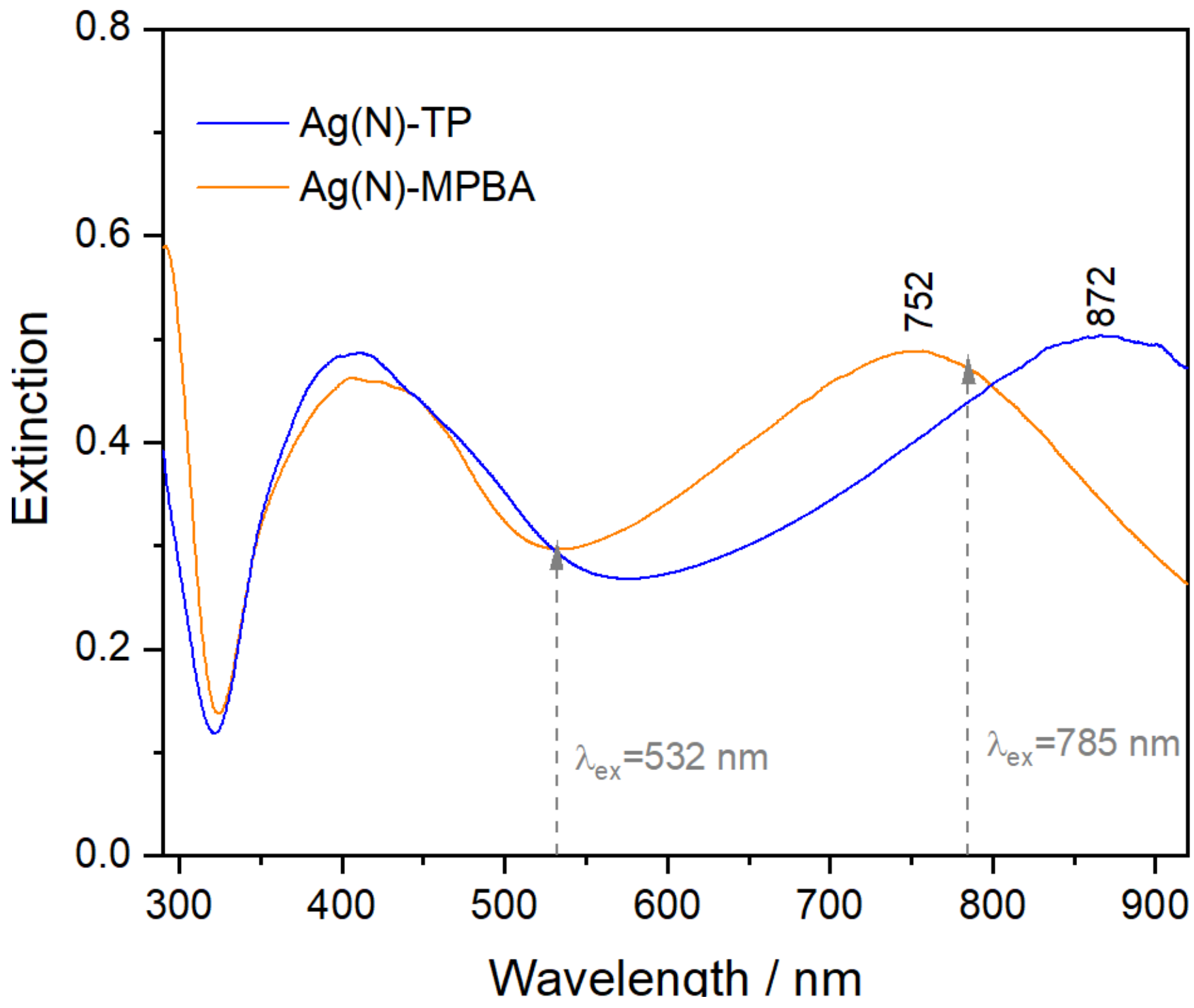

**Figure S3** – Surface plasmon extinction (SPE) spectra of Ag(N)-TP and Ag(N)-MPBA SERS active systems (measured in 1.0 cm cuvette)

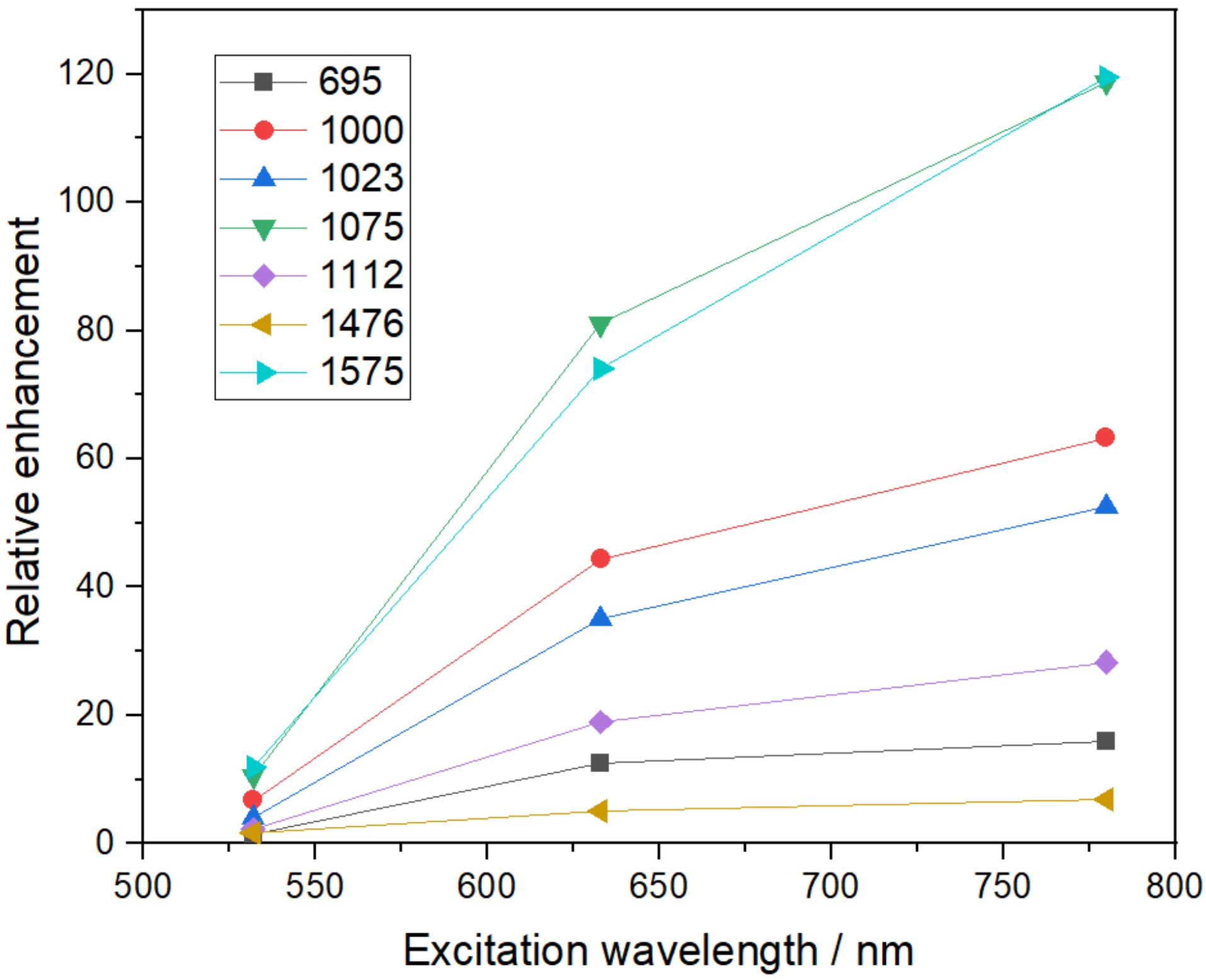

**Figure S4** – SERS excitation profile of selected spectral bands of TP in the Ag(N)-TP system

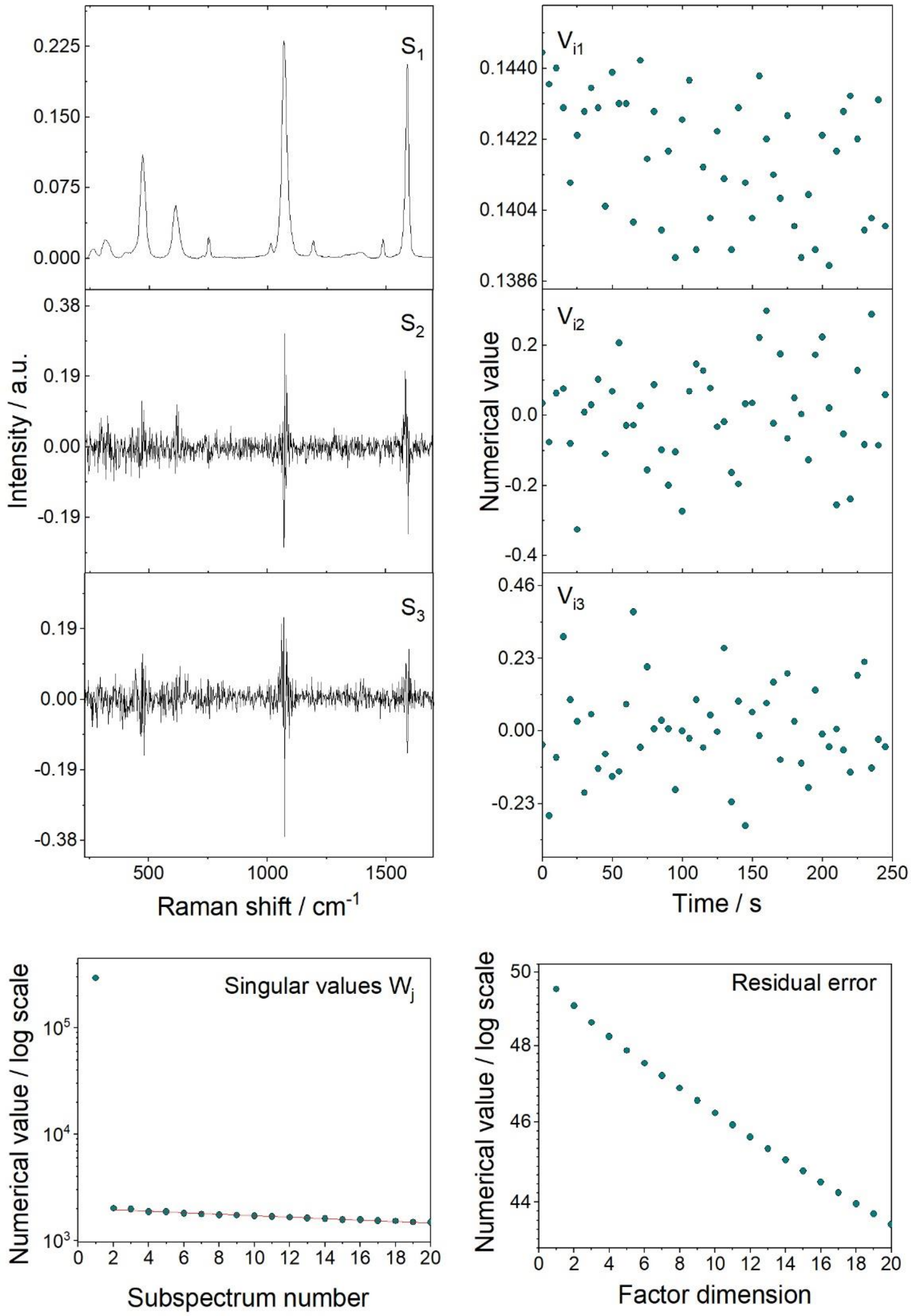

**Figure S5** - Factor analysis of the SERS spectral time series of the Ag(N)-MPBA system in a 1.0-cm cuvette at 5 s accumulation and 785 nm excitation

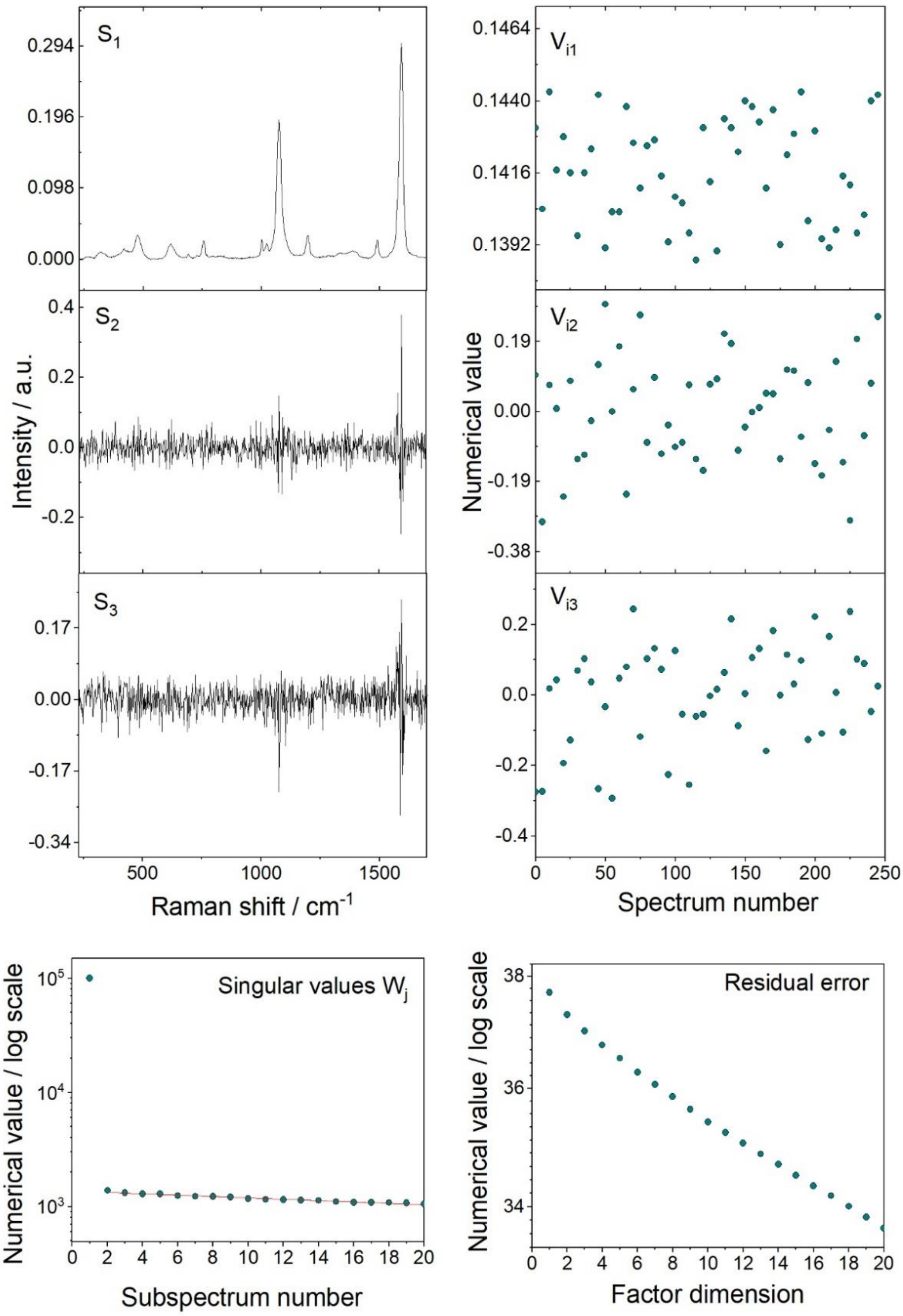

**Figure S6** – Factor analysis of the SERS spectral time series of the Ag(N)-MPBA system in a 1.0-cm cuvette at 5 s accumulation and 532 nm excitation.

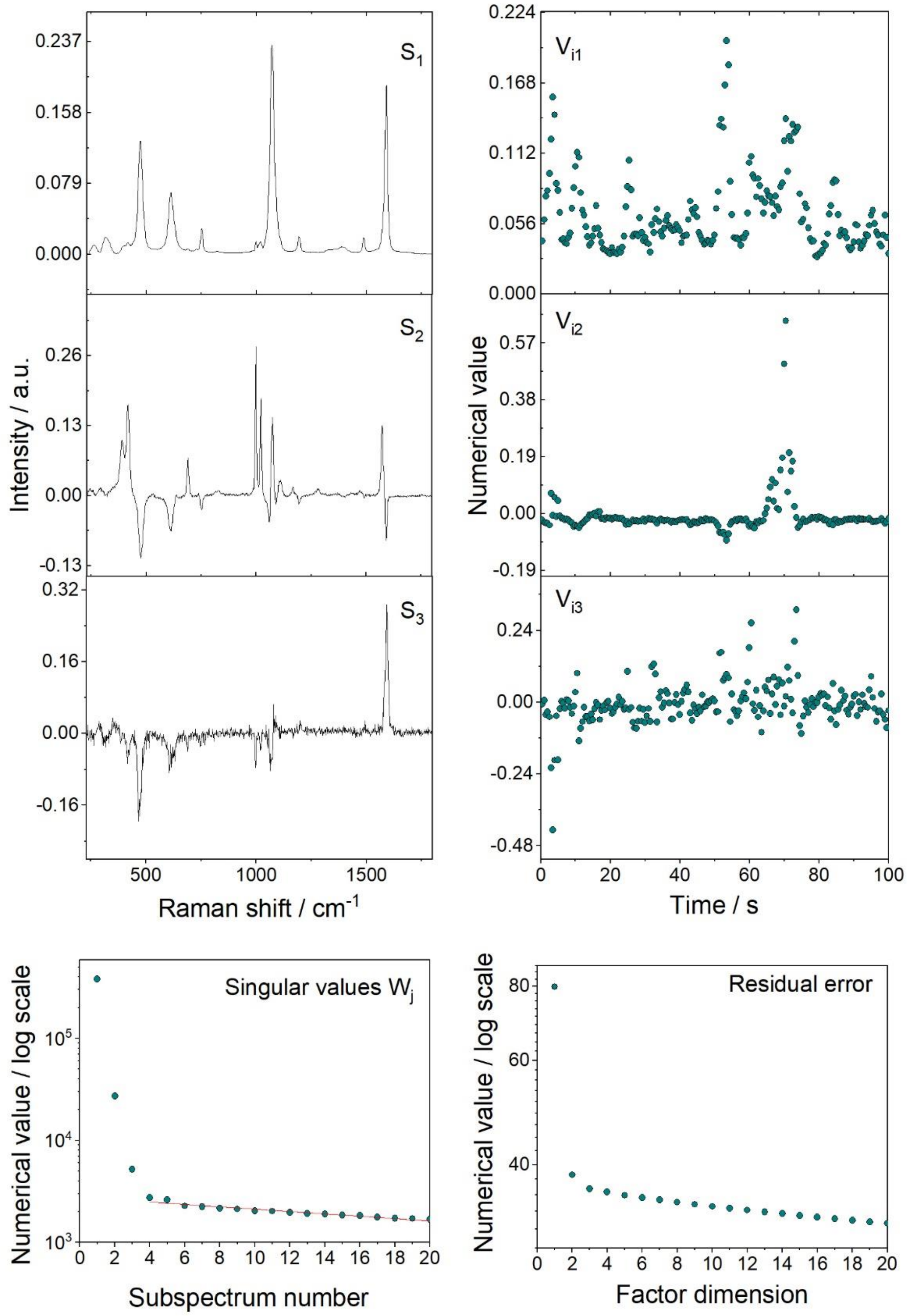

**Figure S7** – Factor analysis of the SERS spectral time series of the Ag(N)-MPBA system in a 0.2-cm cuvette at 0.5 s accumulation and 785 nm excitation

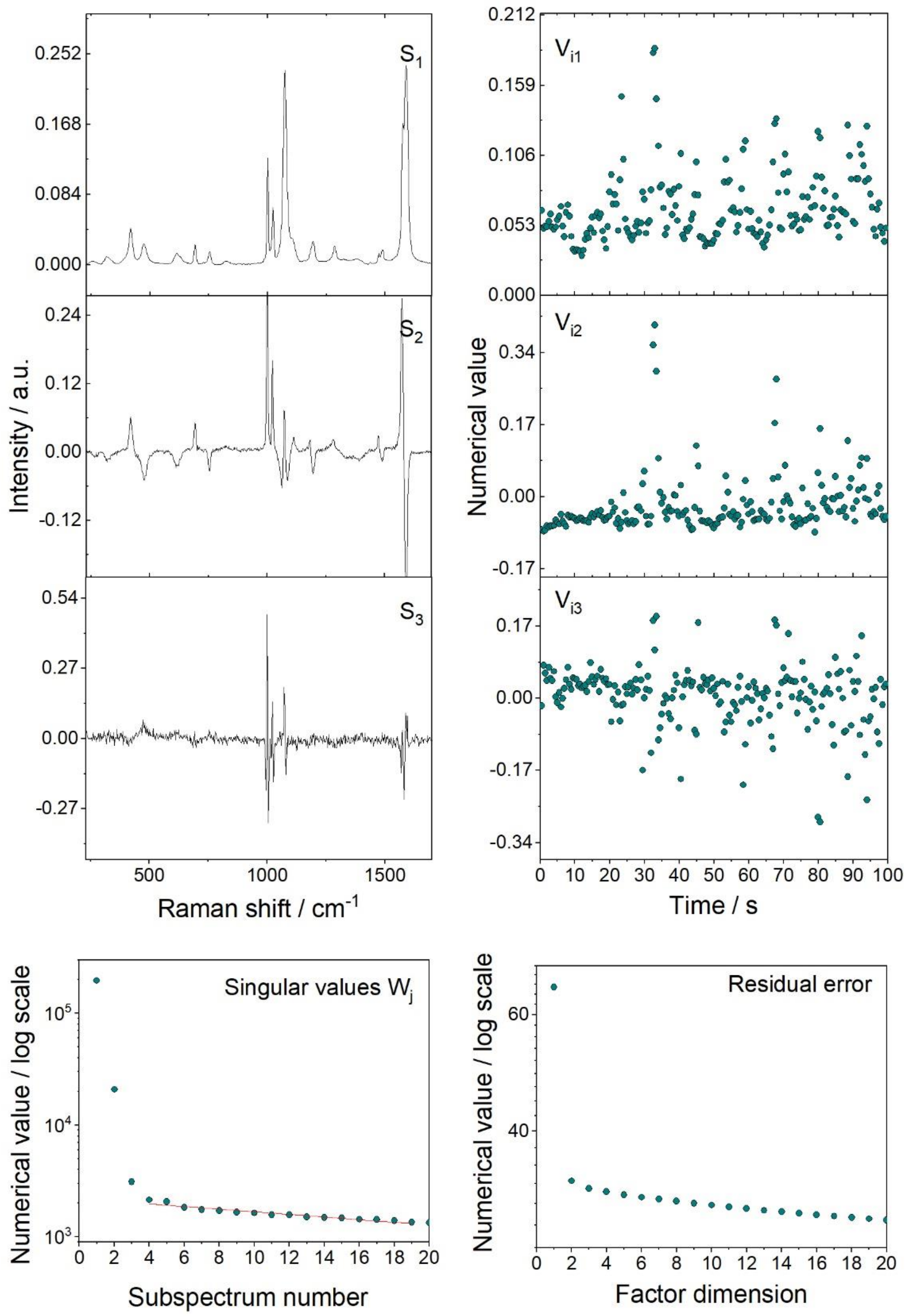

**Figure S8** - Factor analysis of the SERS spectral time series of Ag(N)-MPBA system in a 0.2-cm cuvette, 0.5 s accumulation and 532-nm excitation

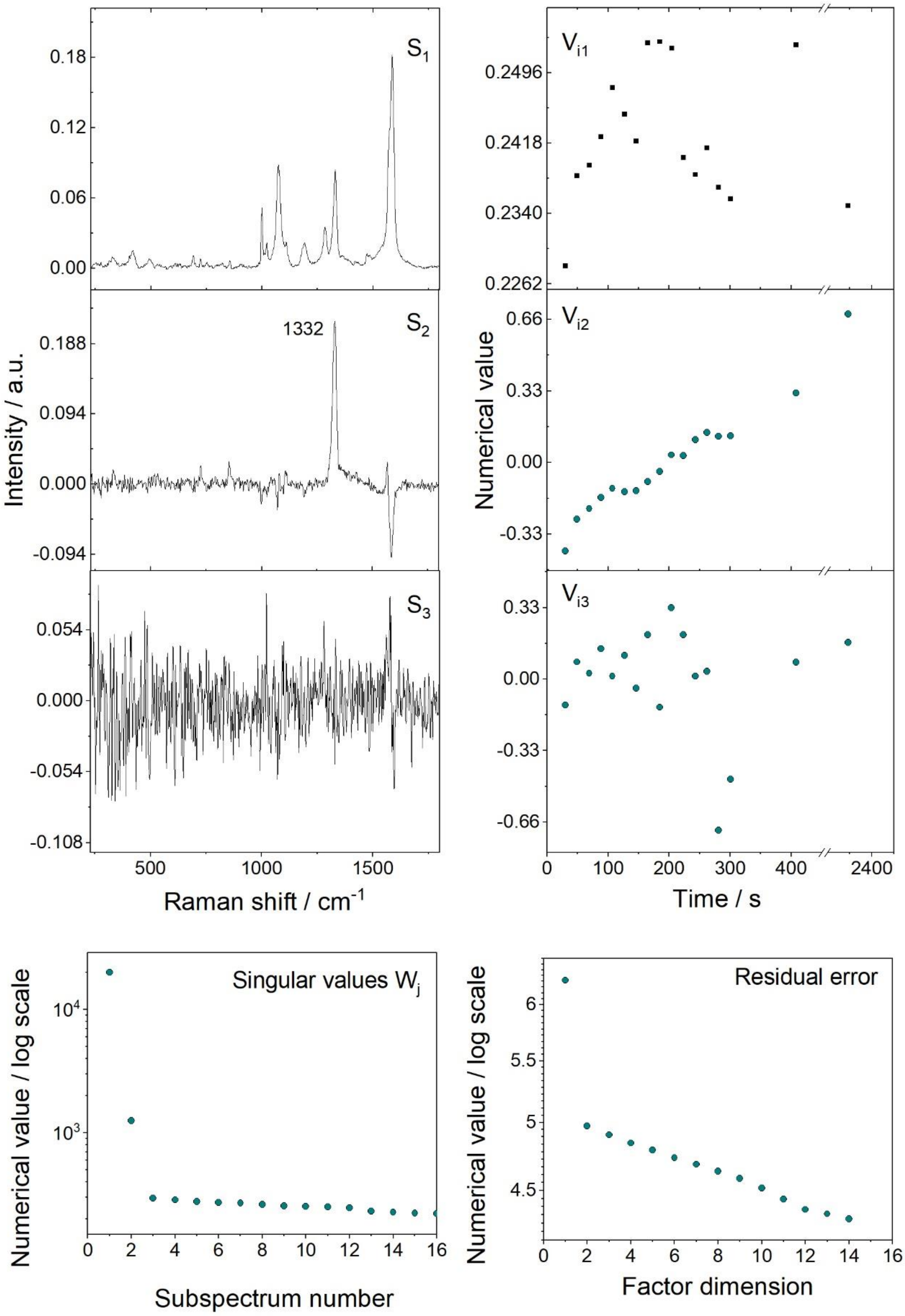

**Figure S9** – Factor analysis of the SERS spectral time series of the Ag(N)-MPBA system excited at 445 nm in a 1.0-cm cuvette at 10 s accumulation.

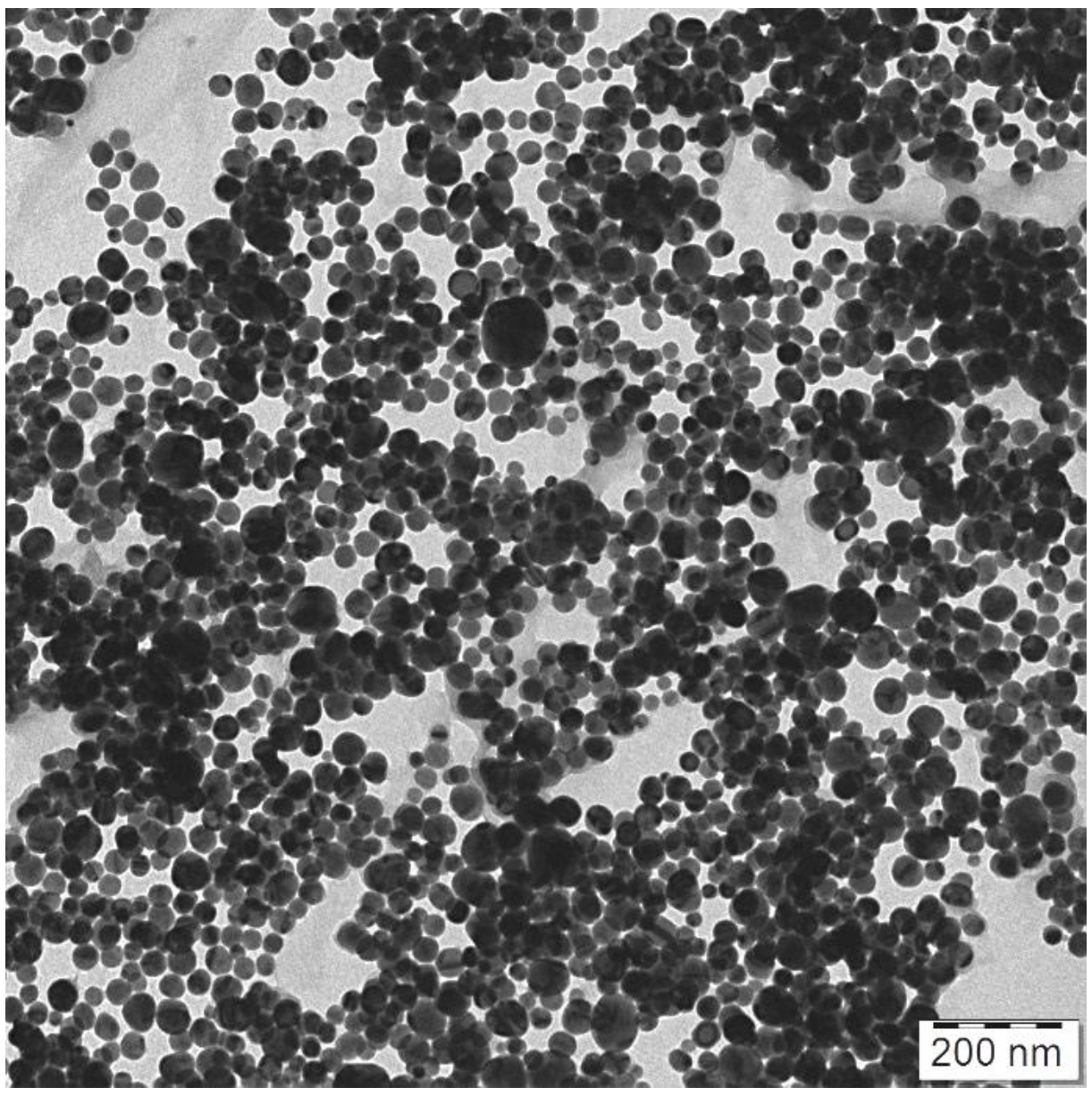

**Figure S10** - Transmission electron microscope (TEM) image of the Ag(N)-MPBA deposited system

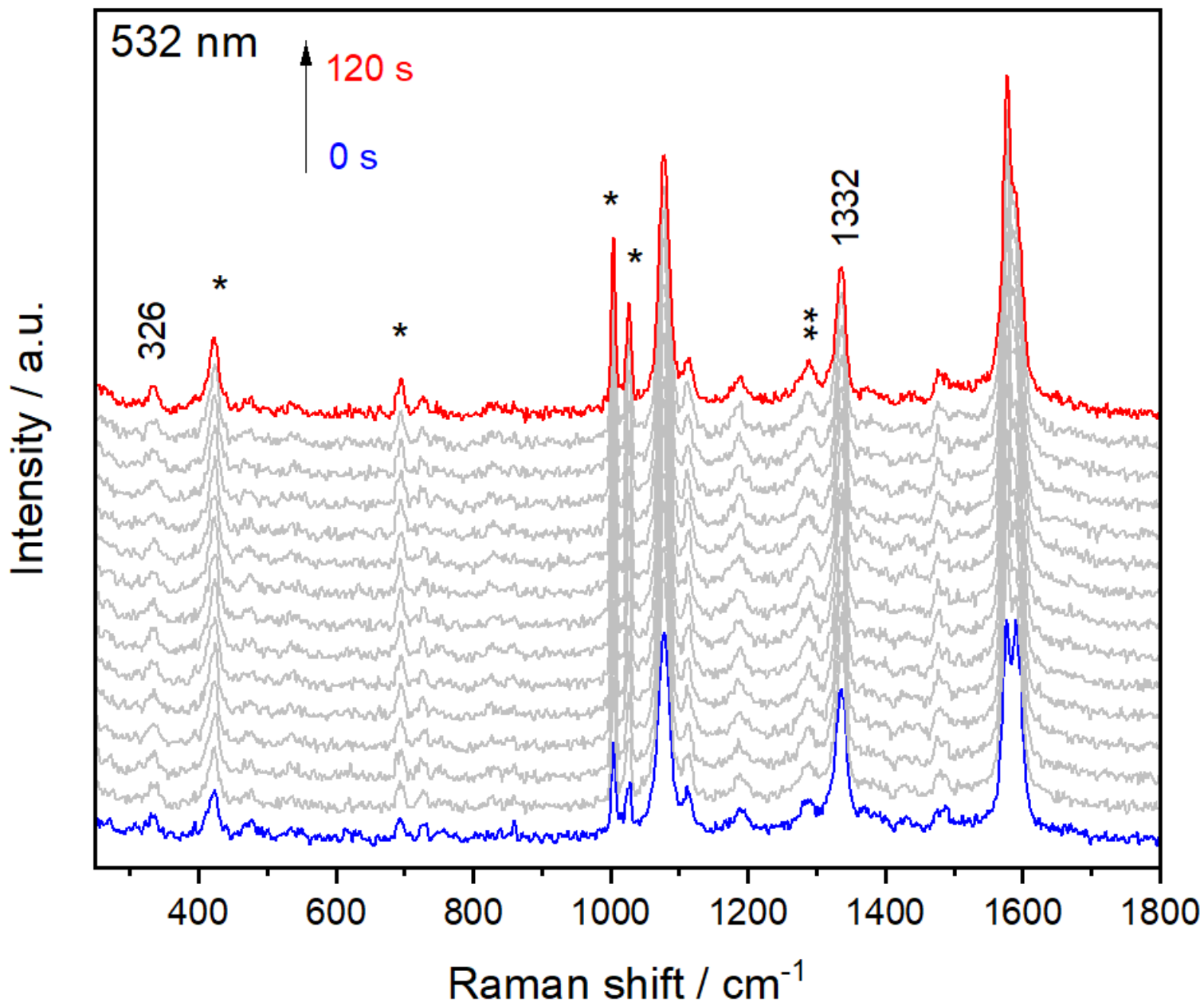

**Figure S11** – Time-evolution SERS spectra of the 2D deposited array of the Ag(N)-MPBA system measured at 532 nm excitation, 0.1 mW laser power and 0.5 s accumulation

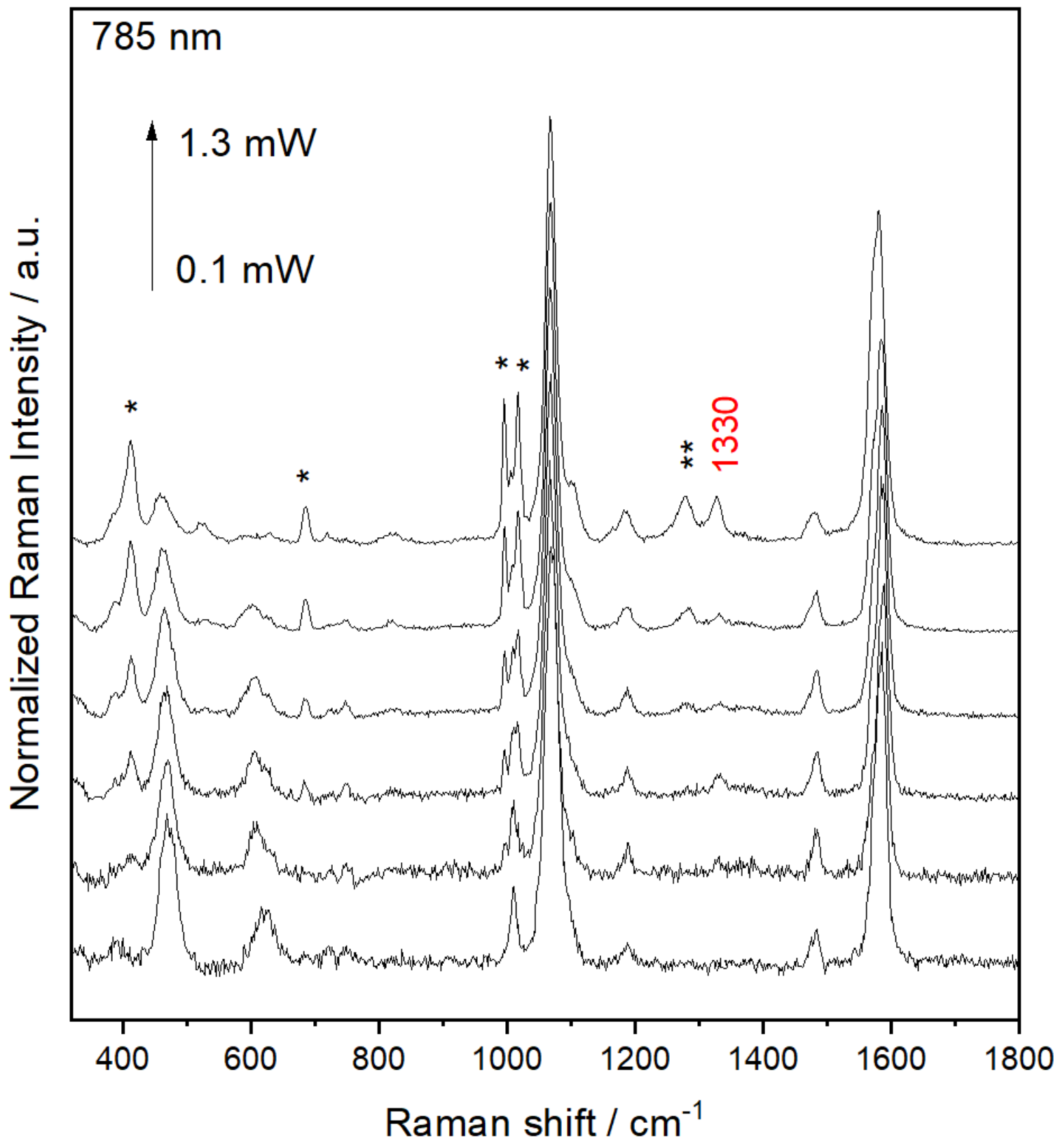

**Figure S12** - 2D deposited array of the Ag(N)-MPBA system measured with a variable laser power (0.1 - 1.3 mW) at 785 nm excitation and 0.5 s accumulation. TP bands are marked with asterisk (*), and BPDT bands are marked with double asterisks (**).

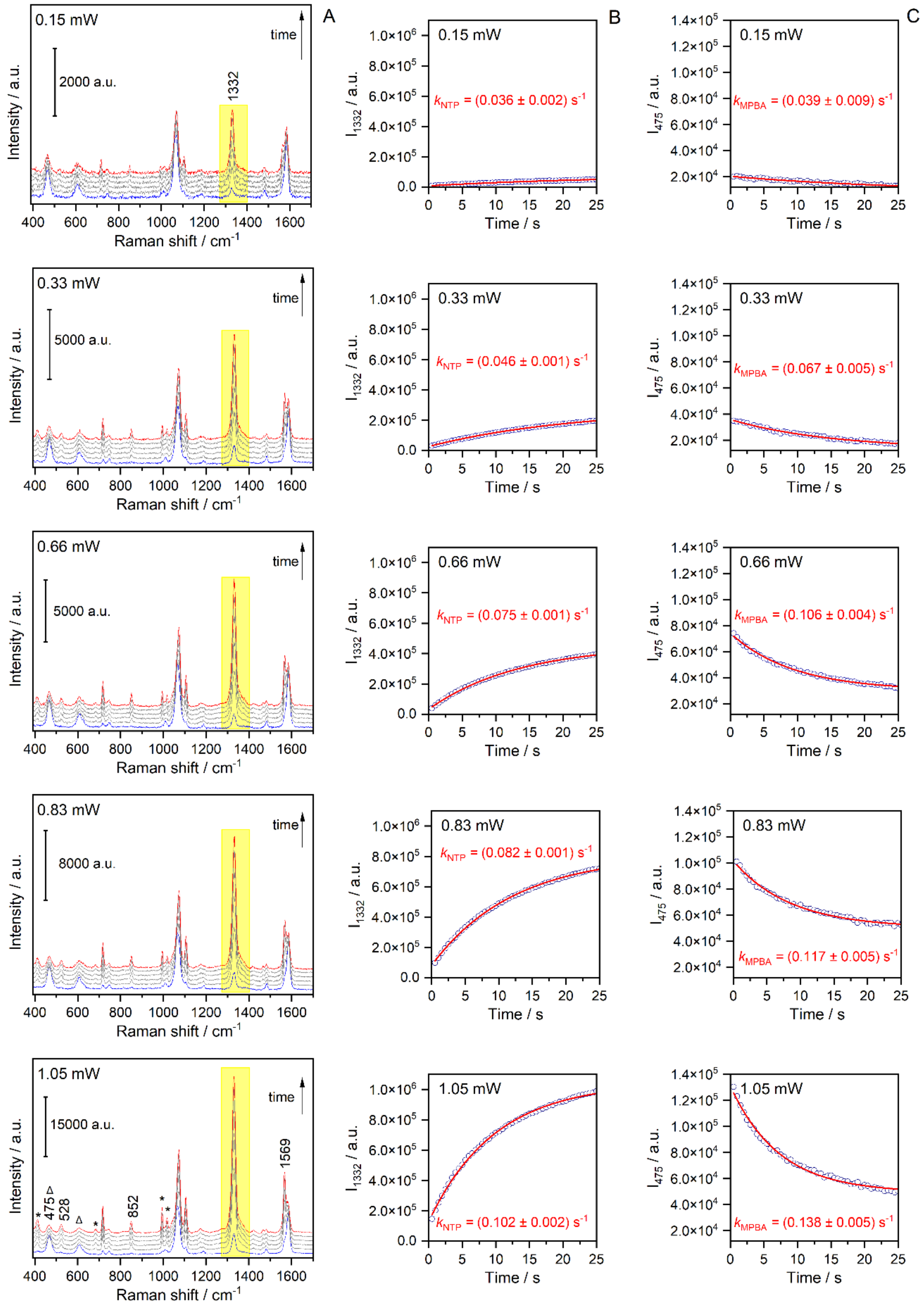

**Figure S13** – A) Time-evolution SERS spectra of a 2D deposited array of the Ag(N)-MPBA system covered with a 0.2 M solution of $NaNO_2$. The laser power ranged from 0.15 to 1.05 mW, as shown in the images, at $\lambda_{ex}$=785 nm and 0.5 s accumulation, plotting every 10$^{th}$ spectrum. B) integral intensity of the 1332 cm$^{-1}$ band as a function of time; C) integral intensity of the 475 cm$^{-1}$ band as a function of time. The red lines in B and C represent the fit of the curve using the mono-exponential function $I = I_0 + A * \exp(-k \cdot t)$. Rate constants $k$ in units s$^{-1}$ are shown in the images. TP bands are marked with an asterisk, and MPBA bands are marked with a triangle.